\documentclass[aps,showpacs,preprintnumbers, superscriptaddress, nofootinbibt,twocolumn]{revtex4}
\usepackage{amsmath}
\usepackage{eurosym}
\usepackage{amssymb}
\usepackage{graphicx}
\usepackage{color}

\def\be{\begin{equation}}
\def\ee{\end{equation}}
\def\bea{\begin{eqnarray}}
\def\eea{\end{eqnarray}}

\begin{document}

\title{Mass bounds for compact spherically symmetric objects in generalized gravity theories}
\author{Piyabut Burikham}
\email{piyabut@gmail.com}
\affiliation{High Energy Physics Theory Group, Department of Physics, Faculty of Science,
Chulalongkorn University, Phyathai Rd., Bangkok 10330, Thailand}
\author{Tiberiu Harko}
\email{t.harko@ucl.ac.uk}
\affiliation{Department of Physics, Babes-Bolyai University, Kogalniceanu Street,
Cluj-Napoca 400084, Romania}
\affiliation{Department of Mathematics, University College London, Gower Street, London
WC1E 6BT, United Kingdom}
\author{Matthew J. Lake}
\email{matthewj@nu.ac.th}
\affiliation{The Institute for Fundamental Study, ``The Tah Poe Academia Institute", \\
Naresuan University, Phitsanulok 65000, Thailand \\
Thailand Center of Excellence in Physics, Ministry of Education, Bangkok
10400, Thailand }
\date{\today }

\begin{abstract}
We derive upper and lower bounds on the mass-radius ratio of stable compact objects in extended gravity theories, in which modifications of the gravitational dynamics via-{\' a}-vis standard general relativity are described by an effective contribution to the matter energy-momentum tensor. Our results include the possibility of a variable coupling between the matter sector and the gravitational field and are valid for a large class of generalized gravity models. The generalized continuity and Tolman-Oppenheimer-Volkoff equations are expressed in terms of the effective mass, density and pressure, given by the bare values plus additional contributions from the total energy-momentum tensor, and general theoretical limits for the maximum and minimum mass-radius ratios are explicitly obtained. As applications of the formalism developed herein, we consider compact bosonic objects, described by scalar-tensor gravitational theories with self-interacting scalar field potentials, and charged compact  objects, respectively. For Higgs type models, we find that these bounds can be expressed in terms of the value of the potential at the surface of the compact object. Minimizing the energy with respect to the radius, we obtain explicit upper and lower bounds on the mass, which admits a Chandrasekhar type representation. For charged compact objects, we consider the effects of the Poincar\'{e} stresses on the equilibrium structure and obtain bounds on the radial and tangential stresses. As a possible astrophysical test of our results, we obtain the general bound on the gravitational redshift for compact objects in extended gravity theories, and explicitly compute the redshift restrictions for objects with nonzero effective surface pressure. General implications of minimum mass bounds for the gravitational stability of fundamental particles and for the existence of holographic duality between bulk and boundary degrees of freedom are also considered.

\end{abstract}

\maketitle

%Sec.1%%%%%%%%%%%%%%%%%%%%%%%%%%%%%%%%%%%%%%%%%%%%%%%%
%%%%%%%%%%%%%%%%%%%%%%%%%%%%%%%%%%%%%%%%%%%%%%%%%%%%
\section{Introduction} \label{sect1}

General Relativity (GR), given by Einstein's field equations, is highly successful at describing gravitational dynamics at the scale of the Solar System. It is a geometric theory that establishes a beautiful relation between the curvature of spacetime and the configuration of matter fields, and a large number of astronomical observations, as well as terrestrial experiments, have confirmed its predictions in various scenarios. These include observations in both the weak gravity regime present at the Solar System level \cite{Shapiro,Will:2001mx,Anderson_etal} and in the strong gravity regime that describes gravitational wave emission from binary systems of spinning compact objects, including black holes, as recently detected by LIGO \cite{Abbott:2016blz,TheLIGOScientific:2016wfe}. However, though fully consistent with the predictions of GR for black holes with masses in the range $36^{+5}_{-4} M_{\odot}$ and $29^{+4}_{-4} M_{\odot}$ \cite{Abbott:2016blz,TheLIGOScientific:2016src}, the LIGO results also remain consistent with modified gravity models (MOG) for smaller black holes with masses of order $M \lesssim 10M_{\odot}$ \cite{Moffat:2016gkd}, leaving a window for alternative gravity theories \cite{Konoplya:2016pmh}. Furthermore, several recent observations suggest that GR may be unable to describe gravitational phenomena at very large scales, comparable to the present day size of the Universe, motivating the study of MOG to describe cosmological dynamics. In this paper, we investigate the implications of MOG theories for the formation and properties of compact objects, observations of which represent another key test of gravitational dynamics.

The two most serious challenges faced by canonical GR are the apparent existence of dark energy and dark matter. A large number of cosmological observations, obtained initially from distant Type Ia Supernovae, have convincingly shown that the Universe is currently undergoing late time accelerated expansion \cite{1n,2n,3n,4n,acc}. The ``standard" explanation for this is based on the assumption of the existence of a mysterious component, called dark energy, which is responsible for the observed characteristics of late time evolution {\it within} GR \cite{PeRa03,Pa03}. In this scenario, a second mysterious component of the Universe, called dark matter, which was initially introduced to explain the flat rotation curves of galaxies, as well as the virial mass discrepancy at the galaxy cluster level, is also required \cite{Sanders,Bertone:2004pz}.

Usually, dark matter is assumed to be non-baryonic and non-relativistic, and can be detected only through its gravitational interactions at the scale of galaxies or clusters, or by observations of the motion of massive hydrogen clouds \cite{dm1}.  However, the particle nature of dark matter remains unknown. Among the most plausible candidates are weakly interacting massive particles, or WIMPs, whose presumed properties place them beyond the standard model of particle physics \cite{dm2}. Due to their massive nature, WIMPS are slow-moving and therefore represent a particle candidate for ``cold dark matter" (CDM).

In the simplest model, able to account for the current observational data, the so-called Cosmological Concordance or $\Lambda$CDM model, dark energy takes the form of a cosmological constant, whose experimental value is determined as $\Lambda = 3 \times 10^{-56}$ $\rm cm^{-2}$ \cite{Ostriker:1995rn,Tegmark:2003ud,Tegmark:2000qy,Hazra:2014hma,Zunckel:2008ti}. Recent evidence obtained from galaxy survey data suggests that GR, in the presence of a cosmological constant, is able to explain redshift-space distortions up to $z \sim 1.4$, when the Universe was approximately 9 billion years old \cite{Okumura:2015lvp}. This represents one of the most stringent tests of GR yet performed, but still leaves room for non-$\Lambda$CDM cosmologies at earlier times. In particular, recent results also suggest that a model with time-varying vacuum energy gives a better fit to existing data than standard concordance cosmology \cite{Gomez-Valent:2016nzf,Fritzsch:2016ewd,Sola:2016jky,Sola:2016vis}, again motivating the study of MOG. %SNIa+BAO+H(z)+LSS+BBN+CMB

Thus, an interesting alternative model of the Universe, able to explain both the galaxy rotation curves and the late time accelerated expansion, contains a mixture of cold dark matter and ``quintessence", represented by a slowly-varying, spatially inhomogeneous energy density \cite{8n}. From a particle physics viewpoint, quintessence can be implemented by assuming the existence of a scalar field $Q$ with a self-interaction potential $V(Q)$. When the potential energy density of the quintessence field is greater than its kinetic energy density, the pressure $p=\dot{Q}^{2}/2-V(Q) $ associated with the quintessence $Q$-field becomes negative, driving cosmological expansion. The properties of quintessential cosmological models have been extensively studied in the literature (for a recent review see \cite{Tsu}). The existence of a scalar field $\phi$, minimally coupled to gravity via a negative kinetic energy, may also explain the recent acceleration of the Universe, since this gives rise to an effective equation of state, $w_{\rm DE}\rho_{\rm DE} c^2 = p_{\rm DE}$, with $w_{\rm DE} < -1$. Here, $\rho_{\rm DE}$ denotes the mass density of the field and $p_{\rm DE}$ denotes the effective pressure. Such fields, known as phantom fields, were proposed in \cite{phan1}.

Hence, scalar fields, either real or complex, may play a fundamental role in the physical processes describing the evolution of our Universe. If so, the possibility that scalar fields can condense to form massive astrophysical objects can not be rejected {\it a priori}. Such objects, called Boson Stars, may arise as solitonic solutions in canonical GR, in which gravity is minimally coupled to a massive, free, complex scalar field \cite{sol1, sol2}. Generally, solitons are mathematical solutions of strongly nonlinear evolution equations describing localized (particle-like) objects with finite energy. Thus, they may be interpreted physically as the ``particles" of the theory under consideration. Nonetheless, it is important to note that, in many ways, solitons differ greatly from the elementary particles of quantum field theories. In particular, they are either dynamical in nature, or have a non-trivial topological structure, which is responsible for their stability \cite{ds1}.

For free fields, the properties of Boson Stars are described by only two parameters (or scales): Newton's constant $G$, which may be expressed equivalently in terms of the Planck mass or length,
\be
m_{\rm Pl} = \sqrt{\frac{\hbar c}{G}} \, , \quad l_{\rm Pl} = \sqrt{\frac{\hbar G}{c^3}} \,,
\ee
respectively, and the scalar field mass $m$, which may be expressed equivalently in terms of the Compton wavelength \cite{sol3}
\be
\lambda_{\rm C} = \frac{\hbar}{mc} = \frac{l_{\rm Pl}m_{\rm Pl}}{m} \, .
\ee
The maximum mass of a Boson Star is inversely proportional to the mass of the field, so that the smaller the scalar field mass, the larger the maximum mass of the star. By including a quartic self-interaction potential, the maximum mass of a Boson Star can be significantly increased, reaching (or even exceeding), the order of magnitude values for neutron stars \cite{sol3,Colpi}. The inclusion of the rotation further increases the upper mass limit \cite{Ryan}. In addition, under certain conditions, matter inside compact general relativistic objects can also form a Bose-Einstein Condensate (BEC). This possibility has been intensively investigated in the literature (see \cite{Glenn} for a detailed discussion of the condensation processes in astrophysics), and  the existence of stars with majority matter content in the form of a BEC cannot be excluded by present observations \cite{Ha1, Ha2}. The matter inside a BEC star obeys a polytropic equation of state with polytropic index $n=1$, and string-like objects composed of polytropic BEC matter, which resemble dark matter filaments, may also have formed in the early Universe \cite{Harko:2014pba}.

 For a class of self-gravitating matter models, with spherically symmetric field configurations, general scaling arguments were developed in \cite{Heus1,Heus2} and applied to both the Einstein-Yang-Mills system and the Einstein sigma model. In these scenarios, the Schwarzschild mass can be expressed as a non-local functional of the matter variables {\it only}. Furthermore, the behaviour of this functional with respect to the scaling transformations yields important physical information about the system. For example, by using scaling properties, one can exclude particle-like solutions in some cases, whereas, for other models, one can obtain virial relations that include gravitational effects. 

In general, a key parameter used to distinguish between different types of compact astrophysical objects, such as white dwarfs, neutron stars and black holes, as well as in determining the outcome of many astrophysical processes, including supernova explosions and the merger of binaries, is the maximum mass. The theoretical values of the maximum mass and radius of a white dwarf/neutron star were derived by Chandrasekhar and Landau, respectively, and are given by \cite{ShTe83}
\be\label{Ch}
M_{\max} \approx \frac{m_{\rm Pl}^3}{m_{\rm B}^2} \, , \quad R_{\max} \approx \frac{\hbar}{mc}\left(\frac{m_{\rm Pl}^2}{m_{\rm B}}\right) \, ,
%M_{\max}\sim \left(\frac{\hbar c}{G}m_{B}^{-4/3}\right) ^{3/2}, \quad R_{\max} \sim \frac{\hbar }{mc}\left( \frac{\hbar c}{Gm_{B}^{2}}\right) ^{1/2},
\ee
where $m_{B}$ is the mass of the baryons, and $m$ is either the electron mass $m_e$ (for white dwarfs) or the neutron mass $m_n$ (for neutron stars). It is important to note that, in the case of white dwarfs, even though the star is supported by electron degeneracy pressure, most of the mass is in the form of baryons. Thus, with the exception of composition-dependent numerical factors, the maximum mass of a degeneracy supported star depends only on fundamental physical constants. For non-rotating neutron stars with finite central density $\rho_c$, an upper bound of approximately $3M_{\odot}$, where $M_{\odot} = 2 \times 10^{33}$ g is the solar mass, has been found \cite{Rh74}. For quark stars, obeying a linear equation of state of the form $p = a\left(\rho c^2-\rho _0\right)$, where $a$ and $\rho _0$ are constants, the maximum mass and radius of the star have been obtained as \cite{HaCh02}
\begin{eqnarray}\label{12*}
M_{\max} &=& \frac{4}{3}\frac{R_{0}^{3}c^{3}}{\left(a+1\right) ^{3/2}G} \frac{1}{\sqrt{\pi G\rho _{0}}} \, ,
\nonumber\\
R_{\max} &=& \frac{R_{0}c}{\sqrt{\pi \left(a+1\right) G\rho _{0}}} \, ,
\end{eqnarray}
where $R_{0}\approx 0.474$. In fact, one of the most fundamental results in GR-based theoretical astrophysics is the existence of a universal maximum mass-radius ratio for a compact spherically symmetric object, proved by Buchdahl \cite{Bu59}:
\begin{eqnarray}\label{Buchdahl}
\frac{2GM}{c^2R} \leq \frac{8}{9} \, .
\end{eqnarray}
This bound has been generalized to account for compact objects in Schwarzschild-de Sitter geometry \cite{ MaDoHa00}, for charged compact objects \cite{MaDoHa01}, and for fluid spheres with anisotropic pressures \cite{BoHa06}. Comparing the quark star limits (\ref{12*}) with the universal bound (\ref{Buchdahl}), we see that $R_0^2/(a+1) \leq 1/3$. Alternative bounds on the mass-radius ratio for both neutral and charged objects, in the presence of dark energy in the form of a cosmological constant, were obtained in \cite{An1,An2,An3, An4} and \cite{div1,div2}, respectively. Buchdahl type inequalities in $D$-dimensional spacetimes were derived in \cite{W1}, for standard GR, and in \cite{W2} for five-dimensional Gauss-Bonnet gravity. The generalization of the Buchdahl limit for $f(R)$ gravity theories was obtained in \cite{Go15}. In such theories, extra-massive stable stars can exist, with surface redshifts larger than 2. Since this represents the maximum possible surface redshift for a stable compact object in GR, this result may provide an observational test for the validity of $f(R)$ type generalized gravity models. In \cite{new1} it was pointed out that the compactness limit of a dense star is also marked by gravitational field energy exterior to star being less than half its gravitational mass.

 A lower bound on the total mass of a static, spherically symmetric (Schwarzschild) black hole, $M\leq \kappa \mathcal{A}/4\pi$, where $\mathcal{A}$ and $\kappa $ denote the area and surface gravity of the horizon, respectively, was derived in \cite{Heus3}, under the requirement that matter fields obeys the dominant energy condition. By applying this result to scalar fields, one can recover the well-known result that the only black hole solution of the spherically symmetric Einstein-Higgs model, with arbitrary non-negative potential, is the Schwarzschild spacetime with constant Higgs field.  A stronger bound for the total mass of a Reissner-Nordstr{\" o}m type black hole, involving the electromagnetic potentials and charges, was also obtained. These estimate provide a simple but powerful tool to prove a ``no-hair'' theorem for matter fields violating the strong energy condition. 

In the Cosmological Concordance model, the equation of state for dark energy is $\rho_{\Lambda} c^2 = -p_{\Lambda}$, where $\rho_{\Lambda}c^2$ and $p_{\Lambda}$ denote the energy density and effective pressure associated with the cosmological constant $\Lambda$. This has important theoretical implications in cosmology and astrophysics, which have been studied intensively in the literature, though its possible effects on the microscopic structure of matter have been less thoroughly investigated. In \cite{min1} it was shown that, in the framework of the classical GR, the presence of a positive cosmological constant  implies the existence of a minimal mass and of a minimal density in nature, such that
\be \label{dens_lim}
\frac{2GM}{c^2}\geq \frac{\Lambda }{6}R^3 \, , \quad \rho = \frac{3M}{4\pi R^3} \geq \rho_{\Lambda} \equiv \frac{\Lambda c^2}{16 \pi G} \, . %\quad (\Lambda \geq 0) \, .
\ee
These results rigorously follow from the generalized Buchdahl inequality in the presence dark energy, described by $\Lambda \geq 0$. The astrophysical and cosmological implications of the existence of a minimum density and mass due to the presence of the cosmological constant were considered in \cite{min2}, where a representation of the cosmological constant in terms of ``classical" fundamental constants was also obtained:
\be \label{Lambda_ident}
\Lambda \approx \frac{\hbar^2G^2m_e^6c^6}{e^{12}} \, .
\ee
Equation (\ref{Lambda_ident}) closely resembles a remarkably prescient result originally obtained by Zel'dovich \cite{Zel'dovich:1968zz,Zel1,Zel2}, was first noticed as numerical coincidence in \cite{Funkhouser:2005hp}, and has been ``derived" using Minimum Length Uncertainty Relations (MLURs) \cite{Hossenfelder:2012jw,Ng:1993jb}, motivated by phenomenological quantum gravity, in \cite{min5}, and by analogy with the Kinchin axioms in information theory in \cite{Khinchin,Beck:2008rd}. In Sect. \ref{sect6} we investigate alternative ways of obtaining this correspondence, including those based on the pioneering work on quantum gravity by Bronstein \cite{Bronstein}, applied to minimum mass constraints obtained for a Universe containing dark energy \cite{min4}.

The bound (\ref{dens_lim}) was generalized for anisotropic objects in \cite{BoHa06}, and for charged objects in \cite{min3}, where it was shown that, for charged fluid spheres with anisotropic internal pressures, in the presence of a positive cosmological constant $\Lambda > 0$, the inequalities
\be
\frac{2GM}{c^2} \geq \frac{\Lambda}{6}R^3+\frac{3}{4}\frac{Q^2}{R} \, , \quad \langle\rho\rangle \geq \frac{c^2\Lambda}{16\pi G} + \frac{9}{8}\frac{Q^2}{R^4} \, ,
\ee
hold in canonical GR, where $\langle\rho\rangle$ is the average density. The generalized Buchdahl inequalities in arbitrary spacetime dimensions with $\Lambda \neq 0$ were obtained in \cite{min4}, by considering both the de Sitter and anti-de Sitter cases. The Jeans instability of barotropic dark energy was also investigated in the framework of a simple $d$-dimensional Newtonian model, both with and without viscous dissipation. The dispersion relation describing the dark energy-matter condensation process was determined, along with estimates of the corresponding Jeans mass (and radius). The minimum and maximum mass-radius ratios of a stable, charged, spherically symmetric compact object in a $D$-dimensional spacetime, in the framework of canonical GR in the presence of dark energy, were obtained in \cite{min5}. By combining the lower bound on the density, in four spacetime dimensions, with ``cubic" MLURs, the limit (\ref{r_e}) was obtained as an upper bound on the charge-mass ratio of {\it any} stable, gravitating, charged quantum mechanical object. In addition, the general minimum charge-mass relation was found to preserve holography between bulk and boundary degrees of freedom in arbitrary dimensions \cite{min5}. These results suggest the existence of a deep connection between gravity, the existence of the dark energy, the stability of fundamental particles and holography.

In order to explain the observed present day acceleration of the Universe, alternatives to ``particle physics" models of dark energy have also been proposed. In such (MOG) theories, dark energy is not represented by specific physical field but, instead, is induced on cosmological scales by intrinsic modifications of the gravitational interaction. Hence, in this case, one can assume that, at large astrophysical and cosmological scales, standard GR is unable to describe the dynamical evolution of the Universe. Many types of modified gravity theory have been proposed in the literature. Some important general classes are: $f(R)$ gravity, in which the gravitational action is an arbitrary function of the Ricci scalar $R$ \cite{Bu701,Bu702,Bu703,Bu704}, $f\left(R,L_{\rm m}\right)$ gravity, in which it is an arbitrary function of the Ricci scalar and the matter Lagrangian $L_{\rm m} $ \cite{fL1,fL2,fL3,fL4}, and $f(R,T)$ gravity theories, in which $T$ denotes the trace of the matter energy-momentum tensor $T^{\mu \nu}$ \cite{fT1,fT2}, the Weyl-Cartan-Weitzenb\"{o}ck (WCW) model \cite{WCW}, hybrid metric-Palatini $f(R,\mathcal{R})$ gravity theories, where $\mathcal{R}$ is the Ricci scalar formed from a metric-independent connection \cite{HM1,HM2}, $f\left(R,T,R_{\mu \nu }T^{\mu \nu}\right)$ type models, where $R_{\mu \nu}$ is the Ricci tensor \cite{Har4}, Eddington-inspired Born-Infeld theory \cite{EIBI}, and $f(\tilde{T},T)$ gravity, in which a coupling between the torsion scalar $\tilde{T}$ and  the trace of the matter energy-momentum tensor is assumed \cite{HT}. For a recent review of the generalized gravity theories with non-minimal curvature-matter coupling, of $f\left(R,L_{m}\right)$ and $f(R,T)$ type, see \cite{Revn}. For a review of hybrid metric-Palatini gravity, see \cite{Revn1}. Current bounds on modified gravity from binary pulsar and cosmological observations were discussed in \cite{new4}, were the potential of future gravitational wave measurements to test the behavior of gravity in the strong-field regime was also emphasized.

Modified gravity models are important because (in principle), they are able to provide a unified theoretical framework for understanding both the late time acceleration of the Universe and the apparent effects of dark matter. In this scenario, dark matter, like dark energy, is not the represented by a physical particle or matter field, but by a fundamental modification of the gravitational interaction.

It is the goal of the present paper to obtain the upper and lower limits for the fundamental physical parameters (mass-radius ratio, maximum and minimum mass, and surface redshift) describing the gravitational structure of compact objects in a large class of  extended gravitational theories. In particular, we consider theories in which modifications of the canonical gravitational dynamics can be described in terms of an effective contribution to the matter energy-momentum tensor. This extra contribution can be of geometric origin, or due to the presence of a ``real" physical field, such as, for example, a scalar field, or the electromagnetic field generated by the presence of charge. Moreover, to ensure our results hold as generally as possible, we include the possibility of a variable coupling between matter and the gravitational field. We derive the generalized continuity and Tolman-Oppenheimer-Volkoff (TOV) equations in terms of the effective mass, density and pressure, given by the sum of the ``bare" values, corresponding to the matter sector, and the additional contributions from the total energy-momentum tensor. In \cite{new2}  a stellar structure formalism was constructed, without adhering to any particular theory of gravity, and which describes in a simple parameterized form the departure from general relativistic compact stars. This post-TOV formalism  is inspired by the parametrized post-Newtonian theory, extended to second post-Newtonian order by adding suitable correction terms to the fully relativistic TOV equations. The post-TOV formalism was extended to deal with the stellar exterior in \cite{new3}, where several potential astrophysical observables were also computed, including the surface redshift, the apparent radius, the Eddington luminosity at infinity, and the orbital frequencies, respectively.

General limits for the maximum and minimum possible mass-radius ratios for gravitationally stable, compact objects are explicitly obtained. As an application of the formalism developed, we consider the case of compact bosonic objects, described by scalar-tensor gravitational theories with self-interacting scalar field potentials, and compact charged objects, respectively. For the self-interaction potential we adopt a Higgs type expression, with quadratic and quartic terms in the scalar field, and derive the maximum and minimum mass bounds in terms of its surface value. Hence, we propose an expression for the minimum mass of a gravitationally stable particle, which takes a form analogous to the Chandrasekhar mass for white dwarfs/neutron stars. In the case of charged compact objects, we also consider the effects of the Poincar\'{e} stresses on the equilibrium structure, and obtain bounds on the radial and tangential stresses. As a possible astrophysical test of our results, we present the general bound on the gravitational redshift for compact objects in extended gravity theories, which may be of use in the observational detection of deviations from standard GR. The redshift restrictions for objects with nonzero effective surface pressure are explicitly obtained.

This paper is organized as follows. In Sect.~\ref{sect2}, we derive the TOV equation for general extended gravity models, with variable gravitational coupling. The maximum and minimum mass limits for this class of theories are obtained in Sect.~\ref{sect3} and the mass limits for scalar-tensor type modifications of standard GR are discussed in detail in Sect.~\ref{sect4}, in which the scalar field is assumed to be minimally coupled to gravity. The mass limits for compact charged objects are considered in Sect.~\ref{sect5}, where limits on the Poincar\'{e} stresses are derived. Applications of minimum mass limitis to the case of microscopic objects (i.e. fundamental particles) are considered in Sect.~\ref{sect7}. Finally, a brief summary and discuss of our results, including a discussion of the surface redshift as a test of generalized gravity theories, and prospects for future work, is presented in Sect.~\ref{sect7}.

%Sec.2%%%%%%%%%%%%%%%%%%%%%%%%%%%%%%%%%%%%%%%%%%%%%%%%
%%%%%%%%%%%%%%%%%%%%%%%%%%%%%%%%%%%%%%%%%%%%%%%%%%%%
\section{Tolman-Oppenheimer-Volkoff equation in generalized gravity theories} \label{sect2}

In the following analysis, we investigate the mass bounds for compact objects in extended gravitational theories. As a first step in our study, we adopt the following representation for the total
energy-momentum tensor of the general modified gravity model:
\begin{equation}
T_{\mu \nu }^{(\rm tot)}=T_{\mu \nu }^{(\rm m)}+\theta _{\mu \nu } \, ,
\end{equation}
where
\begin{equation}
T_{\mu \nu }^{(\rm m)}=\left(\rho c^{2}+p\right) u_{\mu }u_{\nu }-pg_{\mu \nu} \, ,
\end{equation}
is the energy-momentum tensor of the ordinary matter, whose thermodynamic properties are determined by the mass density $\rho$, and thermodynamic pressure $p$. The four-velocity of the matter fluid $u_{\mu }$, is
normalized so that $u_{\mu }u^{\mu }=1$. The tensor $\theta _{\mu \nu }$ describes the geometric or physical properties of any additional term that may arise due to the presence of extra interactions, such as those generated
by the presence of charge, or other ``physical" fields, or because of the extension of the gravitational model.

In many theoretical extensions of canonical GR, the gravitational coupling is time-, space-, or energy-dependent. We therefore allow for the possibility of a varying, or effective, gravitational coupling $G_{\rm eff}$, which is assumed to
have the general form
\begin{equation} \label{Geff}
G_{\rm eff}=\frac{G_{0}}{G} \, ,
\end{equation}
where $G_{0}$ is the present day gravitational ``constant", and $G$ is a function of the spacetime coordinates. Hence, we investigate a general class of gravitational theories in which the gravitational field equations can
be written in the form
\begin{equation} \label{1}
R_{\mu \nu }-\frac{1}{2}Rg_{\mu \nu }=\frac{8\pi G_{0}}{c^{4}}\left[\frac{1}{G}T_{\mu \nu }^{(\rm m)}+\theta _{\mu \nu }\right] \, ,
\end{equation}

Equivalent scalar-tensor formulations of the type described by Eqs.~(\ref{1}) can be obtained for several modified gravity theories. For the case of the  $f(R)$ gravity, the field equations are given by \cite{Bu702,Revn}
\begin{equation} \label{fieldin}
R_{\mu \nu }-\frac{1}{2}g_{\mu \nu }R=8\pi \frac{G_0}{\phi }T_{\mu \nu}^{(\rm m)}+\theta _{\mu \nu }
\end{equation}
where
\begin{equation}
\theta _{\mu \nu }=-\frac{1}{2}V\left( \phi \right) g_{\mu \nu }+\frac{1}{\phi }\left( \nabla _{\mu }\nabla _{\nu }-g_{\mu \nu }\square \right) \phi \, ,
\end{equation}
with the scalar field satisfying the Klein-Gordon equation
\begin{equation} \label{trace1}
3 \square \phi + 2V(\phi) -\phi \frac{dV}{d\phi }=8\pi G\, T^{(\rm m)} \, .
\end{equation}

In the scalar-tensor representation, the field equations of the $f(R)$ gravity theory can be obtained from a Brans--Dicke type gravitational action, with parameter $\omega =0$, given by
\begin{equation}\label{scalx}
S = \frac{1}{16\pi G}\int \left[ \phi R-V(\phi) +L_{m}\right] \sqrt{-g} \; d^{4}x \, ,
\end{equation}
where $V(\phi)$ is the self-interaction potential of the scalar field. The $f\left(R,L_m\right)$ theory with linear curvature matter-coupling can be reformulated as a scalar-tensor theory, which an be derived from the action \cite{Revn1}
\begin{equation} \label{300}
S = \int d^4x \sqrt{-g} \left[ \frac{\psi R }{2} -V(\psi) \, + U(\psi) {\cal L}_m \right],
\end{equation}
where
\begin{eqnarray}
V(\psi) &=& \frac{\phi(\psi) f_1' \left[ \phi (\psi ) \right] - f_1\left[ \phi( \psi ) \right] }{2} \, , \label{400}\\
U( \psi) & =& 1+\lambda f_2\left[ \phi( \psi ) \right] \, ,
\label{500}
\end{eqnarray}
with $f_1$ and $f_2$ arbitrary functions, and $\lambda $ a coupling constant. The so-called hybrid metric-Palatini theory \cite{HM1,Revn1} belongs to the class of the algebraic family of scalar-tensor theories, and can be derived from the action
\bea \label{eq:S_scalar2}
S &=& \frac{1}{2\kappa^2}\int {\rm d}^4 x \sqrt{-g} \Bigg[ (Q_{\rm A}+\phi)R +\frac{3}{2\phi}\partial_\mu \phi \partial^\mu \phi
- %\nonumber\\&&
V(\phi)\Bigg] \nonumber\\ &+& S_m \,,
\eea
where $Q_{\rm A}$ is a constant. The corresponding gravitational and scalar field equations are given by
\begin{eqnarray}\label{einstein_phi}
%&&
\left(Q_{\rm A} + \phi\right)G_{\mu\nu} &=& \kappa^2T_{\mu\nu}^{(\rm m)} + \nabla_\mu\nabla_\nu\phi\nabla_\alpha \nabla^\alpha \phi\/g_{\mu\nu}
\nonumber\\
&-& \frac{3}{2\phi}\nabla_\mu\phi \nabla_\nu\phi + \frac{3}{4\phi}\nabla_\lambda\phi\nabla^\lambda\phi g_{\mu \nu}
\nonumber\\
&-& \frac{1}{2}Vg_{\mu\nu} \, ,
\end{eqnarray}
\begin{equation}\label{eq:evol-phi}
-\nabla_\mu \nabla^\mu \phi+\frac{1}{2\phi}\partial_\mu \phi \partial^\mu\phi+\frac{\phi[2V-(1+\phi)V_\phi]} {3}=\frac{\phi\kappa^2}{3}T^{(\rm m)}.
\end{equation}

In all these gravitational theories the total energy-momentum tensor satisfies the conservation equation
\begin{equation}
\nabla _{\mu }\left[\frac{1}{G}T_{\nu }^{(\rm m)\mu }+\theta _{\nu }^{\mu }\right] =0 \, ,  \label{2}
\end{equation}
which is a direct consequence of the gravitational field equation (\ref{1}).

%Sec.2.1%%%%%%%%%%%%%%%%%%%%%%%%%%%%%%%%%%%%%%%%%%%%%%%
\subsection{The Tolman-Oppenheimer-Volkoff equation} \label{sect2.1}

In the following, we assume a static spherically symmetric spacetime geometry, in which the interior metric inside a massive fluid sphere takes the standard form:
\begin{equation}  \label{metr}
ds^2 = e^{\nu(r)}d(ct)^2 - e^{\lambda(r)}dr^2 - r^2\left(d\theta ^2+\sin^2\theta d\phi ^2\right),
\end{equation}
where $\nu $ and $\lambda $ are functions of the radial coordinate $r$, and the coordinate domains are $0 \leq r < \infty$, $0 \leq \theta \leq \pi$ and $0 \leq \phi < 2\pi $. In the comoving reference frame with $u^{\mu}=\left(e^{\nu /2},0,0,0\right)$, the components of the matter energy-momentum tensor are given by $T^{(\rm m)\mu}_{\nu}=\mathrm{diag}\left(\rho c^2,-p,-p,-p\right)$.

For the metric given by Eq.~(\ref{metr}), the gravitational field equations
become \cite{LaLi}
\begin{equation}
-\frac{1}{r^{2}}\frac{d}{dr}\left(re^{-\lambda }\right) +\frac{1}{r^{2}} = \frac{8\pi G_{0}}{c^{4}}\left( \frac{\rho c^{2}}{G}+\theta _{0}^{0}\right) ,
\label{00}
\end{equation}%
\begin{equation}
-e^{-\lambda }\left( \frac{\nu ^{\prime }}{r}+\frac{1}{r^{2}}\right) +\frac{1}{r^{2}}=\frac{8\pi G_{0}}{c^{4}}\left( -\frac{p}{G}+\theta _{1}^{1}\right) ,
\label{rr}
\end{equation}%
\begin{eqnarray}
&-&\frac{1}{2}e^{-\lambda }\left[ \nu ^{\prime \prime } + \frac{\nu^{\prime2}}{2}+\frac{\nu ^{\prime } - \lambda^{\prime }}{r}-\frac{\nu^{\prime}\lambda^{\prime }}{2}\right]
\notag \\
&=&\frac{8\pi G_{0}}{c^{4}}\left( -\frac{p}{G}+\theta _{2}^{2}\right) = \frac{8\pi G_{0}}{c^{4}}\left( -\frac{p}{G}+\theta _{3}^{3}\right) ,
\end{eqnarray}%
where a prime denotes the derivative with respect to the radial coordinate $r$. In the following we will restrict our attention to isotropic models. We therefore require that the tensor $\theta _{\mu}^{\nu}$ satisfies the
condition $\theta _2^2=\theta _3^3$, and the gravitational coupling function $G_{\rm eff}$, is assumed to be a function of only the radial coordinate, so that $G=G(r)$ in Eq. (\ref{Geff}).

The conservation of the effective energy-momentum tensor may be written as
\begin{equation}
\left(\nabla _{\mu }\frac{1}{G}\right) T_{\nu }^{(\rm m)\mu } + \frac{1}{G}\nabla_{\mu }T_{\nu }^{(\rm m)\mu }+\nabla _{\mu }\theta _{\nu }^{\mu} = 0 \, ,
\end{equation}
or, equivalently,
\begin{eqnarray}  \label{3}
&&\left(\nabla _{\mu }\frac{1}{G}\right) T_{\nu }^{(\rm m)\mu } + \frac{1}{G}\Bigg[ \frac{\partial }{\partial x^{\mu }}\ln \sqrt{-g}T_{\nu }^{(\rm m)\mu } + \frac{\partial }{\partial x^{\mu }}T_{\nu }^{(\rm m)\mu}
\notag \\
&-& \frac{1}{2}\frac{\partial g_{\alpha \beta }}{\partial x^{\mu }}T^{(\rm m)\alpha \beta }\Bigg] +\frac{\partial }{\partial x^{\mu }}\ln \sqrt{-g}\theta _{\nu }^{\mu }+\frac{\partial }{\partial x^{\mu }}\theta _{\nu }^{\mu}
\notag \\
&-& \frac{1}{2}\frac{\partial g_{\alpha \beta }}{\partial x^{\mu }}\theta^{\alpha \beta }=0 \, .
\end{eqnarray}

For a static, spherically symmetric system, Eq.~(\ref{3}) gives the condition
\begin{eqnarray}
\frac{G^{\prime }}{G^{2}}p &+& \frac{1}{G}\left[-\frac{1}{2}\left(\rho c^{2}+p\right)\nu^{\prime }-p^{\prime }\right] +\frac{d\theta _{1}^{1}}{dr}
 \notag \\
&+&\frac{1}{2}\left( \theta _{1}^{1}-\theta _{0}^{0}\right) \nu ^{\prime } + \frac{2}{r}\left( \theta _{1}^{1}-\theta _{2}^{2}\right) =0 \, ,
\end{eqnarray}
from which we immediately obtain
\begin{equation} \label{nuprime}
\nu^{\prime} = -\frac{2\frac{d}{dr}\left( \frac{p}{G}-\theta _{1}^{1}\right) - \frac{4}{r}\left( \theta _{1}^{1}-\theta _{2}^{2}\right) }{\frac{\rho c^{2}}{G}+\theta _{0}^{0}+\frac{p}{G}-\theta _{1}^{1}} \, .
\end{equation}
Equation (\ref{00}) can be integrated immediately to give
\begin{equation}
e^{-\lambda }(r)=1-\frac{2G_{0}m_{\rm eff}(r)}{c^{2}r} \, ,
\end{equation}
where
\begin{equation} \label{mass}
m_{\rm eff}(r)=4\pi \int_{0}^{r}\left[ \frac{\rho \left( r^{\prime }\right) }{G\left( r^{\prime }\right) }+\frac{\theta _{0}^{0}}{c^{2}}\right] r^{\prime 2}dr^{\prime } \, .
\end{equation}
Equation (\ref{rr}) yields
\begin{equation}
\nu^{\prime} = \frac{2G_{0}}{c^{2}}\frac{\frac{4\pi }{c^{2}}\left(\frac{p}{G} - \theta _{1}^{1}\right) r^{3} + m_{\rm eff}}{r^{2}\left[ 1-\frac{2G_{0}m_{\rm eff}}{c^{2}r}\right]} \, ,
\end{equation}
which, together with Eq.~(\ref{nuprime}), gives the generalized TOV equation for modified gravity theories with space-dependent gravitational coupling as
\begin{widetext}
\begin{equation}
\frac{d}{dr}\left(\frac{p}{G} - \theta _{1}^{1}\right) =-\frac{G_{0}}{c^{2}}\frac{\left(\frac{\rho c^{2}}{G}+\theta _{0}^{0}+\frac{p}{G}
-\theta_{1}^{1}\right) \left[\frac{4\pi }{c^{2}}\left(\frac{p}{G}
-\theta_{1}^{1}\right) r^{3}+m_{\rm eff}\right] }{r^{2}\left[1-\frac{2G_{0}m_{\rm eff}}{c^{2}r}\right]} - \frac{2}{r}\left(\theta _{1}^{1} - \theta _{2}^{2}\right) \, .
\label{TOV}
\end{equation}
\end{widetext}
Equation (\ref{TOV}) can be formulated in a compact form if we introduce the effective energy density $\rho_{\rm eff}c^{2}$ and the effective pressure $p_{\rm eff}$, defined as
\begin{equation}
\rho _{\rm eff}c^{2} = \frac{\rho c^{2}}{G} + \theta _{0}^{0} \, , \quad p_{\rm eff}=\frac{p}{G} - \theta _{1}^{1} \, .
\end{equation}

The TOV equation can then be reformulated in terms of the effective quantities in the form
\begin{eqnarray}
\frac{dp_{\rm eff}}{dr} &=& -\frac{G_{0}}{c^{2}}\frac{\left(\rho_{\rm eff}c^{2}+p_{\rm eff}\right) \left(\frac{4\pi }{c^{2}}p_{\rm eff}r^{3}+m_{\rm eff}\right) }{r^{2}\left[ 1-\frac{2G_{0}m_{\rm eff}}{c^{2}r}\right] }
\notag \\
&-&\frac{2}{r}\left( \theta _{1}^{1}-\theta _{2}^{2}\right) \, ,
\end{eqnarray}
while $\nu^{\prime }$ can be expressed as
\begin{equation}
\nu^{\prime } = \frac{2G_{0}}{c^{2}}\frac{\frac{4\pi }{c^{2}}p_{\rm eff}r^{3}+m_{\rm eff}}{r^{2}\left[ 1-\frac{2G_{0}m_{\rm eff}}{c^{2}r}\right] }
= \frac{2G_{0}}{c^{2}}\frac{\frac{4\pi }{c^{2}}p_{\rm eff}r^{3} + m_{\rm eff}}{r^{2}e^{-\lambda }} \, .  \label{nu1}
\end{equation}
For the effective mass, we obtain the continuity equation
\begin{equation}
\frac{dm_{\rm eff}}{dr}=4\pi \rho_{\rm eff}r^{2} \, .
\end{equation}
Finally, subtracting Eqs.~(\ref{00}) and (\ref{rr}) gives the important relation
\begin{equation}\label{sum}
\nu^{\prime} + \lambda^{\prime} - \frac{8\pi G_{0}}{c^{4}}\frac{\left(\rho_{\rm eff}c^{2} + p_{\rm eff}\right) r}{e^{-\lambda }} = 0 \, .
\end{equation}

%Sec.3%%%%%%%%%%%%%%%%%%%%%%%%%%%%%%%%%%%%%%%%%%%%%%%%
%%%%%%%%%%%%%%%%%%%%%%%%%%%%%%%%%%%%%%%%%%%%%%%%%%%%
\section{The Buchdahl and minimum mass limits for compact objects in extended gravitational theories} \label{sect3}

By multiplying Eq. (\ref{nu1}) with $e^{\nu /2+\lambda /2}/r$ we obtain the
equation
\begin{equation}
e^{\lambda /2}\frac{1}{r}\frac{d}{dr}e^{\nu /2}=\frac{G_{0}}{c^{2}}\left(\frac{4\pi }{c^{2}}p_{\rm eff}+\frac{m_{\rm eff}}{r^{3}}\right) e^{\nu /2+\lambda/2} \, .
\end{equation}
Taking the derivative of the above equation, we then have
\begin{eqnarray}
&&\frac{d}{dr}\left(e^{\lambda /2}\frac{1}{r}\frac{d}{dr}e^{\nu /2}\right) = \frac{G_{0}}{c^{2}}e^{\nu /2+\lambda /2}\Bigg[\frac{4\pi }{c^{2}}\frac{d}{dr}p_{\rm eff}
\nonumber\\
&+&\frac{d}{dr}\frac{m_{\rm eff}}{r^{3}}+\left( \frac{4\pi }{c^{2}}p_{\rm eff}+\frac{m_{\rm eff}}{r^{3}}\right) \frac{\nu^{\prime } + \lambda^{\prime}}{2}\Bigg] = \frac{G_{0}}{c^{2}}e^{\nu /2+\lambda /2}
\nonumber\\
&\times& \Bigg[ \frac{d}{dr}\frac{m_{\rm eff}}{r^{3}}+\left( \frac{\nu^{\prime } + \lambda^{\prime}}{2} - \frac{4\pi G_{0}}{c^{4}}\frac{\left(\rho_{\rm eff}c^{2} + p_{\rm eff}\right) r}{e^{-\lambda }}\right)
\nonumber\\
&\times& \left(\frac{4\pi }{c^{2}}p_{\rm eff}+\frac{m_{\rm eff}}{r^{3}}\right) - \frac{8\pi}{c^{2}r}\left(\theta_{1}^{1} - \theta _{2}^{2}\right)\Bigg] \, .
\end{eqnarray}

Hence, with the use of Eq.~(\ref{sum}), and by denoting $y(r)=e^{-\lambda (r)/2}$, $\zeta \left(r\right) = e^{\nu (r)/2}$ and $\Delta =\left(G_0/c^4\right)\left(\theta _2^2-\theta _1^1\right)$, we obtain the following identity
\begin{equation}\label{id}
\frac{y}{r}\frac{d}{dr}\left [\frac{y}{r}\frac{d\zeta}{dr}\right ] = \frac{\zeta }{r}\left[\frac{d}{dr}\frac{m_{\rm eff}(r)}{r^{3}}+\frac{8\pi \Delta }{r}\right ] \, .
\end{equation}
The function $\zeta$ satisfies the condition $\zeta = e^{\nu /2}>0,\forall r\in \lbrack 0,R\rbrack$, where $R$ is the vacuum boundary of the compact object. In the following, we assume that inside a compact object, the condition
\begin{equation}\label{cond}
\frac{d}{dr}\frac{m_{\rm eff}(r)}{r^{3}} < 0 \, ,
\end{equation}
representing a monotonic decrease in mass density as a function of radial distance, holds independently of both the gravitational theory and the equation of state governing the matter. Beginning with Eqs.~(\ref{id}) and (\ref{cond}), we can now derive now the maximum and minimum mass limits for compact objects in generalized gravity theories. In the following analysis, we rescale the effective mass so that
%\begin{equation}\label{rescale}
%G_0m_{\rm eff}/c^2\rightarrow m_{\rm eff} \, ,
%\end{equation}
$G_0m_{\rm eff}/c^2\rightarrow m_{\rm eff}$, for the sake of notational simplicity.

%Sec.3.1%%%%%%%%%%%%%%%%%%%%%%%%%%%%%%%%%%%%%%%%%%%%%%%
\subsection{The Buchdahl limit} \label{sect3.1}

We start our derivation of the maximum mass limit by defining the new function
\begin{equation}  \label{10}
\eta(r) = 8\pi \int _{0}^{r} \frac{r'}{y(r')} \left \{\int_{0}^{r'} \frac{\Delta(r'')}{y(r'')} \frac{\zeta(r'')}{r''} dr''
\right \}dr' \, .
\end{equation}
Next, denoting
\begin{equation}
\Psi =\zeta -\eta \, ,
\end{equation}
and introducing the new independent variable
\begin{equation}
\xi =\int _{0}^{r} \frac{r'}{y(r')} dr' \, ,
\end{equation}
we obtain the condition
\begin{equation}  \label{12}
\frac{d^{2}\Psi }{d\xi ^{2}}<0 \, ,\quad \forall r\in \left [0,R\right] \, ,
\end{equation}
from Eq.~(\ref{id}). This is a fundamental result that holds for all compact objects in generalized gravity theories. Using the mean value theorem, it follows that \cite{Stra}
\begin{equation}
\frac{d\Psi }{d\xi }\leq \frac{\Psi \left (\xi \right ) - \Psi(0)}{\xi} \, ,
\end{equation}
and, by taking into account that $\Psi (0)>0$, we obtain the inequality
\begin{equation}
\Psi^{-1}\frac{d\Psi }{d\xi }\leq \frac{1}{\xi } \, .
\end{equation}
In terms of our original variables, Eq~(\ref{12}) may be written as
\begin{multline}  \label{15n}
\frac{y(r)}{r} \left(\frac{1}{2}\frac{d\nu }{dr}e^{\nu (r)/2} - 8\pi\frac{r}{y(r)} \int _{0}^{r}\frac{\Delta(r') e^{\nu(r')/2}}{y(r')r'} dr' \right)
\\
\leq \frac{e^{\nu (r)/2}-8\pi \int _{0}^{r}\frac{r'}{y(r')} \left( \int _{0}^{r'} \frac{\Delta \left(r''\right )e^{\nu(r'')/2}}{y(r'') r''} dr'' \right) dr'}{\int_{0}^{r} \frac{r'}{y(r')} dr'} \, .
\end{multline}

Since, according to our basic assumption for stable compact objects (\ref{cond}), $m_{\rm eff}/r^{3}$ does not increase outwards, it follows that the condition
\begin{equation}
\frac{m_{\rm eff}(r^{\prime })}{r^{\prime}}\geq \frac{m_{\rm eff}(r)}{r}\left (\frac{r^{\prime }}{r}\right )^{2} \, , \quad \forall r^{\prime }\leq r \, ,
\end{equation}
is satisfied at all points inside the compact object \cite{Stra}. We also {\it assume} that the function $\Delta (r)\geq 0,\ \forall r\in \left[0,R\right]$, describing the effects of modified gravity inside the compact object, satisfies the condition
\begin{eqnarray}  \label{16}
&& \frac{\Delta (r^{\prime \prime })e^{\frac{\nu \left (r^{\prime \prime}\right )}{2}}}{r^{\prime \prime }}\geq \frac{\Delta (r^{\prime })e^{\frac{\nu \left (r^{\prime }\right )}{2}}}{r^{\prime }}\geq \frac{\Delta (r)e^{\frac{\nu (r)}{2}}}{r} \, ,
\nonumber\\
&& \forall r^{\prime \prime} \leq r^{\prime} \leq r \, .
\end{eqnarray}
Physically, this condition means that $\Delta $ is a monotonically decreasing function of the radial coordinate $r$. Therefore, we can evaluate the denominator in the right-hand side of Eq.~(\ref{15n}) as follows,
\begin{eqnarray}  \label{17}
&&\int _{0}^{r} \frac{r'}{y(r')}dr^{\prime }\geq
\int _{0}^{r}r^{\prime }\left [1-\frac{2m_{\rm eff}(r)}{r^{3}}r^{\prime 2}\right ]^{-1/2}dr^{\prime }
\nonumber\\
&=&\frac{r^{3}}{2m_{\rm eff}(r)}\left(1-y(r)\right).
\end{eqnarray}
The second term in the bracket of the left-hand side of Eq.~(\ref{15n}) can be estimated as follows:
\begin{widetext}
\bea  \label{19}
\int _{0}^{r} \frac{\Delta \left (r'\right )e^{\nu(r')/2}}{y(r')r'}dr' \geq \frac{\Delta(r)e^{\nu (r)/2}}{r} \int _{0}^{r}\left [1-\frac{2m_{\rm eff}(r)}{r^{3}}r^{\prime 2}\right ]^{-1/2}dr^{\prime }
= \Delta(r)e^{\nu (r)/2}\left [\frac{2m_{\rm eff}(r)}{r}\right]^{-1/2} \arcsin\left (\sqrt{\frac{2m_{\rm eff}(r)}{r}}
\right) \, . \nonumber\\
\eea
\end{widetext}

For the second term in the numerator of the right-hand side of Eq.~(\ref{15n}), we find
\begin{widetext}
\begin{eqnarray} \label{20}
&&\int _{0}^{r}\frac{r'}{y(r')} \left\{\int_{0}^{r'}\frac{\Delta(r'')e^{\nu(r'')/2}}{y(r'')r''}dr''\right \}dr'
\geq \int_{0}^{r}r^{\prime 2}\frac{\Delta \left (r^{\prime}\right )e^{\nu \left (y(r')r^{\prime }\right )/2}}{r^{\prime }}
\left[\frac{2m_{\rm eff}(r^{\prime })}{r^{\prime}}\right ]^{-1/2} \arcsin\left(\sqrt{\frac{2m_{\rm eff}(r^{\prime })}{r^{\prime }}}\right)dr^{\prime}
\nonumber\\
&&\geq \frac{\Delta (r)e^{\nu (r)/2}}{r}\int_{0}^{r}r^{\prime 2} \left [1-\frac{2m_{\rm eff}(r)}{r^{3}}r^{\prime 2}\biggl/ \frac{2m_{\rm eff}(r)}{r^{3}}r^{\prime 2}\right ]^{-1/2} \arcsin\left[\sqrt{\frac{2m_{\rm eff}(r)}{r^{3}}}r^{\prime}\right] dr^{\prime}
\nonumber\\
&&=\Delta (r)e^{\nu (r)/2}r^{2}\left [\frac{2m_{\rm eff}(r)}{r}\right ]^{-3/2}\left \{\sqrt{\frac{2m_{\rm eff}(r)}{r}}- y(r) \arcsin\left [\sqrt{\frac{2m_{\rm eff}(r)}{r}}\right ]\right \} \, .
\end{eqnarray}
\end{widetext}
Note that, in order to obtain Eq.~(\ref{20}), we have also used the monotonic increase property of the function $\arcsin(x)/x$ for $x\in \left[0,1\right ]$. Using Eqs.~(\ref{17})-(\ref{20}), Eq.~(\ref{15n}) becomes
\begin{widetext}
\bea  \label{21}
\left \{1-\left [1-\frac{2m_{\rm eff}(r)}{r}\right]^{1/2}\right \}\frac{m_{\rm eff}(r)+4\pi r^{3}p_{\rm eff}(r)}{r^{3}\sqrt{1-\frac{2m_{\rm eff}(r)}{r}}}
\leq \frac{2m_{\rm eff}(r)}{r^{3}}+8\pi \Delta (r)\left\{\frac{\arcsin\left [\sqrt{\frac{2m_{\rm eff}(r)}{r}}\right ]}{\sqrt{\frac{2m_{\rm eff}(r)}{r}}}-1\right \}.
\eea
\end{widetext}
Equation (\ref{21}) is valid for all points inside the compact object and does not depend on the sign of $\Delta $.

As a simple consistency check, we consider first the case $\Delta =0$ and $p_{\rm eff}=p$. By evaluating (\ref{21}) for $r=R$, denoting the total mass of the star by $M$, and assuming that the pressure vanishes at the star's surface, $p(R)=0$, we obtain
\begin{equation}
\frac{1}{\sqrt{1-\frac{2M}{R}}}\leq 2\left [1-\left (1-\frac{2M}{R}\right )^{\frac{1}{2}}\right ]^{-1} \, .
\end{equation}
From the above condition, we immediately obtain the well-known result for canonical GR, the Buchdahl inequality (\ref{Buchdahl}) \cite{Bu701,Stra}. By introducing the mean density of the compact object as $\langle \rho _{\rm eff}\rangle (r)=m_{\rm eff}(r)/r^3$, and denoting
\be \label{f(r)*}
f(r) = 4\pi \frac{\Delta (r)}{\langle \rho_{\rm eff}\rangle (r)}\left\{\frac{\arcsin\left [\sqrt{\frac{2m_{\rm eff}(r)}{r}}\right ]}{\sqrt{\frac{2m_{\rm eff}(r)}{r}}} -1\right \} \, ,
\ee
and
\be
w_{\rm eff}(r)=\frac{p_{\rm eff}}{\langle \rho_{\rm eff}\rangle (r)} \, ,
\ee
respectively, we obtain the generalized Buchdahl inequality for extended gravitational theories as
\be\label{58}
\frac{2m_{\rm eff}(r)}{r}\leq 1-\left[1+\frac{2\left(1+f(r)\right)}{1+4\pi w_{\rm eff}(r)}\right]^{-2} \, .
\ee
For $f=0$ and $w_{\rm eff}=0$, we again recover the standard Buchdahl inequality for GR from the above relation.

%Sec.3.2%%%%%%%%%%%%%%%%%%%%%%%%%%%%%%%%%%%%%%%%%%%%%%%
\subsection{The minimum mass of a compact object in extended gravity theories} \label{sect3.2}

On the vacuum boundary of the compact object, defined by the condition $r=R$, Eq.~(\ref{21}) takes the equivalent form
\begin{equation}\label{an1}
\sqrt{1-\frac{2m_{\rm eff}(R)}{R}} \geq \left[ 1+\frac{2\left( 1+f(R)\right) }{1+4\pi w_{\rm eff}(R)}\right]^{-1} \, .
\end{equation}
For small values of the argument $x$, the function $\arcsin(x)/x-1$, which appears in the definition of $f$, can be approximated as $\arcsin(x)/x - 1 \approx x^{2}/6$. Moreover, we denote the total mass of the compact object as
$m_{\rm eff}(R) = M_{\rm eff}$. Hence, at the vacuum boundary of the compact objects, we can approximate the function $f$ as
\be\label{fapprox}
f(R) \approx \frac{4}{3} \pi  \Delta (R) R^2 \, .
\ee
Therefore, Eq.~(\ref{an1}) can be written as
\bea\label{cond1}
&&\sqrt{1-\frac{2M_{\rm eff}}{R}}\geq \frac{3 \left(M_{\rm eff} + 4 \pi  p_{\rm eff} R^3\right)}{12 \pi  p_{\rm eff} R^3 + M_{\rm eff}\left(9+8 \pi  \Delta (R)  R^2\right)} \, .
\nonumber\\
\eea

By introducing a new variable $u = M_{\rm eff}/R\geq 0$ and by denoting
\be
a= 4\pi p_{\rm eff}\left( R\right)R^2 \, , \quad b = 9 + 8\pi  R^{2}\Delta (R) \, ,
\ee
Eq. (\ref{cond1}) takes the form
\begin{equation}\label{cond2}
\sqrt{1-2u} \geq \frac{3(u+a)}{bu+3a} \, ,
\end{equation}
which may be rewritten as
\bea\label{f1cond}
u\left[6 a (3 a-b+3)+u \left(12 a b-b^2+9\right)+2 b^2 u^2\right]\leq 0 \, ,
\eea
or, equivalently,
\be\label{f2cond}
\left(u-u_1\right)\left(u-u_2\right)\leq 0 \, ,
\ee
where
\be\label{66}
u_{1,2}=\frac{-12 a b+b^2-9\pm (b-3) \sqrt{24 a b+\left(b+3\right)^2}}{4 b^2} \, .
\ee
In order for the inequality (\ref{f2cond}) to hold, the conditions $u\geq u_1$, $u\leq u_2$, or $u\leq u_1$, $u\geq u_2$, must be satisfied simultaneously. These conditions imply the existence of a minimum mass-radius ratio for any compact object in modified gravity theory, if $p_{\rm eff}(R)\neq 0$ or $p_{\rm eff}\neq 0$ and $\Delta \neq 0$. In the first order approximation, we obtain
\be\label{66a}
u_1\approx \left(\frac{4}{9}-\frac{a}{6}\right)+ \left(\frac{8 \pi  R^2}{81}+\frac{7}{27} \pi  R^2 a\right)\Delta \, ,
\ee
\be\label{66b}
u_2\approx  \left(-\frac{1}{2}+\frac{1}{3} \pi  R^2 \Delta \right)a \, ,
\ee
so that
\be
u_2\leq \frac{M_{\rm eff}}{R}\leq u_1 \, .
\ee
By assuming that the total effective pressure vanishes at the surface of the compact object, it follows that $a=0$, and we obtain the condition
\be
2 b^2 u-b^2+9\leq 0 \, ,
\ee
or, equivalently, $u\leq \left(b^2-9\right)/2b^2$. This result shows that the presence of a nonzero anisotropic pressure distribution at the surface of the compact object does {\it not} impose a lower bound on the mass-radius ratio.

Assuming, instead, that the parameter $\Delta$, describing the ``direct" effects of the extended gravity theory, vanishes on the surface of the compact object ($\Delta (R)=0$), we obtain $b=9$, so that
\be\label{cond3}
F(u) \equiv 9 u^2 + 2(3 a-2) u+(a-2) a \leq 0 \, .
\ee
The algebraic equation $F(u)=0$ has the non-trivial roots
\begin{eqnarray*}
u_1&=& \frac{1}{9} \left(2-3 a-2\sqrt{ 1+\frac{3}{2}a}\right) \, ,
\nonumber\\
u_2&=& \frac{1}{16} \left(2-3 a+2\sqrt{1+\frac{3}{2} a}\right) \, .
\end{eqnarray*}
By assuming that $3a/2 \ll 1$, we can approximate the roots $u_1$ and $u_2$ by
\be
u_1=-\frac{1}{2}a,u_2=\frac{1}{9}\left(4-\frac{3a}{2}\right) \, ,
\ee
allowing us to reformulate the condition (\ref{cond3}) as
\be\label{cond4}
\left(u+\frac{1}{2}a\right)\left[u-\frac{1}{9}\left(4-\frac{3a}{2}\right)\right]\leq 0 \, .
\ee
By also assuming that the mass-radius ratio of the compact objects satisfies the constraint
\be\label{condn}
u\leq\frac{1}{9}\left(4-\frac{3a}{2}\right) \, ,
\ee
which, for $a=0$, reduces to the standard Buchdahl limit (\ref{Buchdahl}), it follows that the second term in the condition (\ref{condn}) is always negative. Therefore, in order for this condition to be satisfied, the first term must be positive. Hence, we obtain the following bound for the minimum possible mass of a compact object in alternative gravity theories,
\be
u\geq-\frac{1}{2}a \, .
\ee
Since, for realistic physical objects, $u$ must be a positive quantity, it follows that such a minimum mass exists only if $a<0$ or $p_{\rm eff}(R)<0$. We therefore obtain the final lower bound for the mass-radius ratio $M_{\rm eff}^{(\rm min)}/R$ for a massive compact object in extended gravity theories as
\be\label{minmass}
\frac{M_{\rm eff}^{(\rm min)}}{R} \geq 2\pi \left|p_{\rm eff}\right| R^2 \, .
\ee
It is interesting to note, here, that the existence of a {\it minimum} mass-radius ratio is the direct consequence of the presence of a dominant negative pressure on the objects' vacuum boundary. On the other hand, by assuming that $a$ is small and can be neglected, Eq.~(\ref{condn}) gives the restriction $M_{\rm eff}/R \gtrsim 4/9 = 0.444$, the standard Buchdahl limit from canonical GR (\ref{Buchdahl}). A small value of $a$, for example $a = -0.20$, gives the upper limit $M_{\rm eff}/R \lesssim 0.4777$, which shows that the presence of negative pressure can significantly increase the maximum mass-radius ratio of compact objects in generalized gravity theories, as compared to their general relativistic counterparts.

When the external pressure and density satisfy the conditions $\rho_{\rm eff}(r>R) < 0$ and $p_{\rm eff}(r>R) = w\rho_{\rm eff}$, respectively, where $w = \rm const.$, for example, for a spacetime filled with dark energy with negative energy density, such as a negative cosmological constant $\Lambda < 0$, the bulk spacetime is an asymptotically deformed Anti de Sitter (AdS) space. From the viewpoint of holographic duality, the reduction of the maximum mass-radius ratio for positive $a$ (i.e. for negative $w$ and negligible $\Delta$) implies a lower (higher) deconfinement phase transition temperature of the dual gauge matter living on the boundary for $R >(<) \sqrt{3/\Lambda}$, with $\Lambda =8\pi G \rho_{\rm eff}/c^{2}$. It is therefore interesting to explore the physical interpretations of $p_{\rm eff}$ and $w$ from the viewpoint of the dual gauge theory.

For sufficiently large $\Delta>0$, the maximum mass-radius ratio given by Eqn.~(\ref{66a}) will increase if
\begin{equation}
\Delta >\frac{27a}{2\pi R^{2}(8+21 a)} \, .
\end{equation}
Again, for $R >(<) \sqrt{3/\Lambda}$, this corresponds to an increase (decrease) in the phase transition temperature for the deconfinement of the dual gauge matter on the boundary. The {\it anisotropic stress} $\Delta$ can be thought of as a contribution from the bulk ``charge'' and is proportional to the square of the electric charge, $Q^2$, in the electromagnetic case \cite{min5}. In the typical holographic duality ``dictionary", used to ``translate" between differing interpretations of physical quantities in the the bulk and boundary spacetimes, bulk charge is dual to the number density of the gauge matter on the boundary. This suggests that, in our model, $\sqrt{\Delta}$ is also dual to the number density of the boundary gauge matter.

The minimum mass-radius ratio can be interpreted as the dual of the minimum density of the boundary gauge matter {\it before} it vaporizes into a ``hadron'' gas phase \cite{min5}. Since $\Delta$ is always non-negative, Eqn.~(\ref{66b}) implies that positive $a$ increases the minimum mass-radius ratio, which is dual to the higher critical density for the liquid-gas phase transition in the gauge theory picture.

%Sec.4%%%%%%%%%%%%%%%%%%%%%%%%%%%%%%%%%%%%%%%%%%%%%%%%
%%%%%%%%%%%%%%%%%%%%%%%%%%%%%%%%%%%%%%%%%%%%%%%%%%%%
\section{The upper and lower mass limits for bosonic objects}\label{sect4}

 In its simplest theoretical form, we can define a bosonic object as a self-gravitating configuration of a {\it complex massive scalar field $\Psi$}, described by the Lagrangian \cite{Colpi}
\bea\label{act0}
S_{\rm BO} &=& \int\Bigg(-\frac{c^4}{8\pi G_0}R+\frac{1}{2}\nabla _{\mu}\Psi \nabla ^{\mu}\Psi ^{*}\nonumber\\
&& - \frac{m^2}{2}\left|\Psi \right|^2+\frac{1}{4}\lambda \left|\Psi\right|^4 \Bigg)\sqrt{-g}d^4x ,
\eea
where $m$ is the mass of the field and $\lambda $ is a constant. In order for the gravitational field equations to admit a solution under the condition of static spherical symmetry, they must be satisfied by a time-harmonic scalar field ansatz of the form
\be\label{rep}
\Psi \left(\vec{x}\right)=\psi (r)e^{i\omega t},
 \ee
where $\psi (r)$ {\it is a real valued radial amplitude function} and with $\omega $ is the angular frequency eigenvalue of the bosonic object \cite{Colpi}. Using this representation of the scalar field, explicit boson star models can be constructed. It is interesting to note that, because of the compact object's self-gravity, the ground state of the bosonic star is not a zero-energy state \cite{Colpi}. For large values of the parameter $\Lambda =\lambda m_{Pl}^2/4\pi m^2$, the boson star can be described by an effective equation of state of the form \cite{Colpi}
 \be
 p(\rho)=\frac{4\rho _0}{9}\left[\left(1+\frac{3}{4}\frac{\rho }{\rho _0}\right)^{1/2}-1\right]^2,
 \ee
where $\rho _0=m^4/4\lambda$. Hence, in the following analysis, we consider models of bosonic objects that can be constructed from the general action
\be\label{act1}
S_{BO}=\int\Bigg(-\frac{c^4}{8\pi G_0}R+\frac{1}{2}\nabla _{\mu}\psi \nabla ^{\mu}\psi -V(\psi)+L_m\Bigg)\sqrt{-g}d^4x,
%\, ,
\ee
where $\psi $ is a {\it real wave function}, related to the complex scalar wave function $\Psi $ through Eq.~(\ref{rep}), $V(\psi)$ is the self-interaction potential of the scalar field, and $L_m$ is the Lagrangian density of the ordinary matter. Due to the representation (\ref{rep}) of the {\it complex} scalar field, bosonic objects corresponding to the action (\ref{act1}) can always exist, since the existence of conserved Noether charge associated with the U(1) symmetry stabilizes the field configuration. From a physical point of view, in this approach, we neglect the possible variation of the gravitational ``constant" inside compact general relativistic objects. Such an approximation is justified since, according to present day observations and experiments,  we expect that any significant changes in the magnitude of the gravitational coupling should take place over large time or distance intervals. Thus, such a variation of $G$ would have a minimal impact on the internal structure of general relativistic stars.

In the following we will first adopt an approximate description of the compact general relativistic bosonic objects, in which we  ignore the presence of the metric potential $g^{tt}$ in the expression  of $V(\psi)$, originating in the harmonic time dependence
 contribution, $V(\psi)\sim g^{tt} \psi^2$. However, this is a reasonable approach, which should work well as long as the metric tensor component
 $-g_{tt}=e^{\nu}$ is not very different from one, and it does not have strong
 variations inside the compact object (which is indeed the case for most of the boson stars).

Moreover, in Subsection~\ref{sect4.3}, bosonic configurations described by the {\it real} scalar Higgs potential are also explored. Such Higgs-type configurations have finite energy and can again be represented in the form $\Psi(\vec{x},t)=\psi(\vec{x})e^{-i\omega t}$.  The time-dependent part of the Higgs field stabilizes the field configuration and, therefore, such configurations can form stable compact objects.

\subsection{Effective mass, density and pressure for scalar field models minimally coupled to gravity} \label{sect4.1}

The energy-momentum tensor for a scalar field minimally coupled to gravity is
\bea\label{ente}
T^{\mu}_{\nu}&=&\left(\rho c^2+p\right)u_{\nu } u^{\mu }-\delta_{\nu}^{\mu} p+ \nabla_{\nu}\psi\nabla^{\mu}\psi
\nonumber\\
&-&\delta_{\nu}^{\mu}\left[\frac{1}{2}\,\nabla_{\mu}\psi\nabla^{\mu}\psi-V(\psi)\right] \, .
\eea
In the static, spherically symmetric metric (\ref{metr}) the gravitational field equations for the scalar field take the form
\be\label{s1}
-e^{-\lambda}\left(\frac{1}{r^2}-\frac{\lambda^\prime}{r}\right)+\frac{1}{r^2} =\frac{8\pi G_0 }{c^4}\left( \rho c^2 +\frac{1}{2}\,e^{-\lambda} \psi^{\prime \;2}+V\right) \, ,
 \ee
\be\label{s2}
-e^{-\lambda}\left(\frac{1}{r^2}+\frac{\nu^\prime}{r}\right)+\frac{1}{r^2} =\frac{8\pi G_0 }{c^4}\left(- p -\frac{1}{2}\,e^{-\lambda} \psi^{\prime \;2}+V\right) \, ,
\ee
\begin{eqnarray}
&-&\frac{1}{2}e^{-\lambda }\left[ \nu ^{\prime \prime }+\frac{\nu ^{\prime2}}{2}+\frac{\nu ^{\prime }-\lambda ^{\prime }}{r}-\frac{\nu ^{\prime}\lambda ^{\prime }}{2}\right]
\notag \\
&=&\frac{8\pi G_{0}}{c^{4}}\left( -p+\frac{1}{2}\,e^{-\lambda} \psi^{\prime \;2}+V\right) \, .
\end{eqnarray}
The variation of the action with respect to the scalar field gives the Klein-Gordon equation as the EOM for $\psi$,
\begin{equation}\label{s4}
\psi^{\prime\prime}+\left[\frac{2}{r}+\frac{1}{2}\left(\nu^\prime-\lambda^\prime\right)\right]\psi^\prime = e^{\lambda}\frac{d V}{d \psi} \, .
\end{equation}

Equation (\ref{s1}) can be rewritten as
\be
\frac{d}{dr}\left(re^{-\lambda}\right) = 1 - \frac{8\pi G_0}{c^4}\left[\frac{r}{2}\left(re^{-\lambda}\right)\psi^{\prime \;2}+\left(\rho c^2+V\right)r^2\right] \ .
\ee
By representing $e^{-\lambda}$ as
\be
e^{-\lambda} = 1 - \frac{2G_0m_{\rm eff}(r)}{c^2r} \, ,
\ee
it follows that the effective mass of the scalar-tensor theory can be obtained as a solution of the differential equation
\be\label{effdens}
\frac{dm_{\rm eff}}{dr}=-\frac{4\pi G_0}{c^4}r\psi^{\prime\;2}m_{\rm eff}+\frac{4\pi}{c^2}\left(\frac{1}{2}\psi^{\prime\;2}+\rho c^2+V\right)r^2 \, .
\ee
The general solution of Eq. (\ref{effdens}) is
\bea
m_{\rm eff}(r) &=& \frac{4\pi}{c^2}e^{-\frac{4\pi G_0}{c^4}\int{r\psi ^{\prime \;2}dr}}\Bigg\{\int e^{\frac{4\pi G_0}{c^4}\int{r\psi ^{\prime \;2}dr}}
\nonumber\\
&\times& \Bigg[\left(\frac{1}{2}\psi^{\prime\;2}+\rho c^2+V\right)r^2\Bigg]dr \Bigg\} \, ,
\eea
where we have set the arbitrary integration constant equal to zero. By denoting
\be
g(r) = e^{-\frac{4\pi G_0}{c^4}\int{r\psi^{\prime \;2}dr}} \, , \quad \frac{dg}{dr} = -\frac{4\pi G_0}{c^4}r\psi ^{\prime \;2}g \, ,
\ee
we obtain
\be
\psi^{\prime\;2} = -\frac{c^4}{4\pi G_0}\frac{1}{rg}\frac{dg}{dr} \, .
\ee
Thus, we can represent the effective mass as
\bea
m_{\rm eff}(r) &=& \frac{4\pi}{c^2}g(r)\int_0^r\Bigg[\frac{c^4}{8\pi G_0}\frac{1}{r^{\prime}}\frac{d}{dr^{\prime}}\frac{1}{g\left(r^{\prime}\right)}
\nonumber\\
&+& \frac{\rho \left(r^{\prime}\right)c^2+V}{g\left(r^{\prime}\right)}\Bigg]r^{\prime \;2}dr^{\prime} \, .
\eea

Equation (\ref{effdens}) can be written as
\begin{equation}
\frac{dm_{\rm eff}}{dr} = 4\pi \rho_{\rm eff}r^{2} \, ,
\end{equation}
where we have introduced the effective density defined as
\bea
\rho_{\rm eff}(r)&=&\frac{1 }{c^{2}}\Bigg\{ \frac{1}{r^{2}}\frac{dg}{dr}\int_{0}^{r}\Bigg[\frac{c^{4}}{8\pi G_{0}}\frac{1}{r^{\prime }}\frac{d}{dr^{\prime }}\frac{1}{g\left(r^{\prime }\right) }
\nonumber\\
&+&\frac{\rho \left(r^{\prime }\right) c^{2}+V}{g\left(r^{\prime}\right) }\Bigg]r^{\prime\;2}dr^{\prime }
\nonumber\\
&+& \frac{c^{4}}{8\pi G_{0}}\frac{1}{r}\frac{d}{dr}\frac{1}{g\left(r\right)} + \frac{\rho \left(r^{\prime }\right) c^{2}+V}{g\left(r\right) }\Bigg\} \, .
\eea
For the effective pressure, we obtain
\bea
p_{\rm eff}(r) &=& p + \frac{1}{2}e^{-\lambda }\psi ^{\prime \;2}-V
\nonumber\\
&=& p - \frac{c^{4}}{8\pi G_{0}}\left(1-\frac{2G_{0}m_{\rm eff}}{c^{2}r}\right) \frac{1}{rg}\frac{dg}{dr} - V
%\, ,
\eea
while for the parameter $\Delta$ we have
\begin{equation}
\Delta(r) = e^{-\lambda }\psi ^{\prime \;2}=-\frac{c^{4}}{4\pi G_{0}}\left(1 - \frac{2G_{0}m_{\rm eff}}{c^{2}r}\right)\frac{1}{rg}\frac{dg}{dr} \, .
\end{equation}

In the regions of the spacetime where the spatial variation of the scalar field potential can be neglected, so that $dV/d\psi \approx 0$, multiplying the Klein-Gordon equation (\ref{s4}) by $\psi^{\prime}$, we obtain the differential equation
\be \label{KG*}
\frac{d}{dr}\psi ^{\prime \;2}+\left[\frac{4}{r}+\left(\nu ^{\prime}-\lambda ^{\prime}\right)\right]\psi ^{\prime \;2}=0 \, .
\ee
The general solution of Eq. (\ref{KG*}) is
\be
\psi ^{\prime \;2} = \frac{\Psi _0^{\prime}}{r^4}e^{\lambda -\nu} \, ,
\ee
where $\Psi _0^{\prime}$ is an arbitrary constant of integration. The parameter $\Delta(r)$ then becomes
\be\label{Delta}
\Delta(r) =\frac{\Psi _0^{\prime}}{r^4}e^{ -\nu} \, .
\ee

%Sec.4.2%%%%%%%%%%%%%%%%%%%%%%%%%%%%%%%%%%%%%%%%%%%%%%%
\subsection{Maximum and minimum masses for bosonic objects} \label{sect4.2}

For minimally coupled, complex, massive scalar fields, the maximum mass of a bosonic object has been found to of the order of the scalar field's Compton wavelength \cite{sol1,sol3}, being given by
\be
M_{\rm BS}^{\rm max}\approx \alpha _{\rm BS}\frac{m_{\rm Pl}^2}{m} = \alpha _{\rm BS}\times 10^{-9}\times \left(\frac{{\rm GeV}}{m}\right)M_{\odot} \, ,
\ee
where $\alpha _{\rm BS}$ is a numerical coefficient of the order of unity. For scalar field masses of the order of those predicted by the Standard Model of particle physics, the maximum mass is very small and the corresponding objects are called mini-Boson Stars. Much higher mass values can be obtained by including the self-interaction of the scalar field. For spherically symmetric Boson Stars, in theories with quartic self-interaction potentials, it was shown in \cite{Colpi} that the maximum mass is of the order
\be
M_{\rm BS}^{\rm max} \approx 0.062\sqrt{\eta}\frac{m_{\rm Pl}^3}{m^2} \approx 0.062 \times \sqrt{\eta}\times \left(\frac{{\rm GeV}}{m}\right)^2M_{\odot} \, ,
\ee
where $\eta >0$ is the self-interaction coupling for the quartic potential $ V\left(\left|\Psi \right|\right) = \eta \left|\Psi \right|^4$.  The inclusion of rotation can further increase the maximum mass of a Boson Star \cite{Ryan}. In the following, we will restrict our analysis to scalar field potentials of the Higgs type,
\be\label{pot}
V(\psi) = -\frac{m^2}{2}\psi ^2+\frac{\eta}{4} \psi^4 \, ,
\ee
where $m^2\geq 0$ and $\eta \geq 0$. Moreover, we may assume that, at the surface of the compact object $r = R$, the Higgs potential reaches its minimum value, so that $dV(\psi)/d\psi|_{r=R} =0$, giving
\be
\left.\psi \right|_{r=R}=\pm \sqrt{\frac{m^2}{\eta}} \, , \quad \left.V(\psi)\right|_{r=R}=-\frac{1}{4}\frac{m^4}{\eta} \, .
\ee
When the Higgs field is nonminimally coupled to gravity, there exists a family of spherically symmetric particle-like solutions to the field equations \cite{new5}. These monopoles are the only globally regular and asymptotically flat distributions with finite energy of the Higgs field around compact objects.

Using the Klein-Gordon equation (\ref{s4}), it is straightforward to show that the conservation of the total energy-momentum tensor gives the following relation for the ordinary matter pressure $p$:
\begin{equation}\label{s5}
\frac{d p}{d r} = -\frac{1}{2}(\rho c^2+p)\frac{d\nu}{d r} \, .
\end{equation}

%Sec.4.2.1%%%%%%%%%%%%%%%%%%%%%%%%%%%%%%%%%%%%%%%%%%%%%%
\subsubsection{The maximum mass of a bosonic object in generalized gravity theories} \label{sect4.2.1}

As a first step in obtaining the mass bounds for bosonic objects we assume that, near their vacuum boundary, the potential $V$ becomes (approximately) constant, $\left.V(\psi)\right|_{r\in \left(R-\epsilon,R\right)}\approx {\rm const.}$, where the scale length $\epsilon$ satisfies the condition $\epsilon /R \ll 1$. In addition, we assume that the thermodynamic pressure of the bosonic matter $p$ either vanishes, or takes a constant (nonzero) surface value, giving $\left.\nu \right|_{r=R} = {\rm const}$. From Eq.~(\ref{Delta}), it follows that, near the surface of the bosonic object, the approximations $\psi^{\prime 2} \propto 1/r^4 \approx 0$ and $\Delta \propto 1/r^4\approx 0$ are valid. It then follows that the total mass of the Boson Star can be defined as
\be
M_{\rm eff} = m_{\rm eff}(R) = \frac{4\pi}{c^2}\int_0^R{\left(\rho c^2+V\right)r^2dr} = M_B + M_{\psi} \, ,
\ee
where $M_B = 4\pi \int_0^R{\rho r^2dr}$ is the baryonic mass, while $M_{\psi}=\left(4\pi/c^2\right)\int_0^R{V\left(\psi\right)r^2dr}$ is the mass of the scalar field. The effective pressure near the objects' surface becomes
\be
p_{\rm eff}(R) = p(R) - \left.V(\psi)\right|_{r=R} \, .
\ee
Hence, with the use of the above assumptions, the generalized Buchdahl inequality (\ref{21}) gives the following expression for the maximum mass of a compact object in scalar-tensor theories with non-minimally coupled scalar fields:
\begin{equation}
\left\{1-\left[ 1-\frac{2M_{\rm eff}}{R}\right]^{1/2}\right\}\frac{M_{\rm eff} + 4\pi R^{3}p_{\rm eff}(R)}{\sqrt{1-\frac{2M_{\rm eff}}{R}}}\leq 2M_{\rm eff} \, .
\label{cs1}
\end{equation}

Equation (\ref{cs1}) can be rewritten as
\begin{equation}
\frac{2M_{\rm eff}}{R} \leq 1 - \left(\frac{M_{\rm eff}/R + 4\pi R^{2}p_{\rm eff}(R)}{3M_{\rm eff}/R + 4\pi R^{2}p_{\rm eff}(R)}\right)^{2} \, ,
\end{equation}
which gives the generalized Buchdahl identity for compact bosonic objects in scalar-tensor gravity in a more familiar form as
\begin{equation}
\frac{2M_{\rm eff}}{R}\leq \frac{4}{9}\left[1 - 6\pi R^{2}p_{\rm eff}(R) + \sqrt{1 + 6\pi R^{2}p_{\rm eff}(R)}\right] \, .
\end{equation}
Assuming that $6\pi R^{2}p_{\rm eff}(R) \ll 1$, and explicitly reintroducing the physical constants for the sake of clarity, we obtain
\be
\frac{2G_0M_{\rm eff}}{c^2R} \leq \frac{8}{9}\left[1-\frac{3\pi G_0}{2c^4}R^2p_{\rm eff}(R)\right]
\ee
in the first order of approximation. By taking into account the explicit expression for the scalar field potential (\ref{pot}), we obtain for the upper bound on the mass-radius ratio for bosonic objects with Higgs type potentials as
\be
\frac{2G_{0}M_{\rm eff}}{c^{2}R} \leq \frac{8}{9}\left\{1+\frac{3\pi G_{0}}{2c^{4}}\left[\frac{m^{2}}{2}\psi^{2}(R) - \frac{\eta}{4} \psi^{4}(R)\right]R^{2}\right\} \, .
\ee
Assuming, in addition, that at the object's surface the Higgs potential has a minimum, we obtain
\be
\frac{2G_{0}M_{\rm eff}}{c^{2}R}\leq \frac{8}{9}\left\{ 1+\frac{3\pi G_{0}}{8c^{4}}\frac{m^4}{\eta}R^{2}\right\} \, .
\ee

Finally, we can estimate the mass of the scalar field contribution as
\bea
M_{\psi}(R) &\approx& \frac{4\pi R^3}{3}\left.V(\psi)\right|_{r=R}
\nonumber\\
&=& \frac{4\pi R^3}{3}\Bigg[ -\frac{m^{2}}{2}\psi^{2}(R) + \frac{\eta}{4} \psi^{4}(R)\Bigg]
\nonumber\\
&\geq& 0 \, .
\eea
Thus, by also assuming that the pressure of the baryonic matter vanishes at the surface of the compact objects we obtain the following restriction on the maximum mass of the ordinary matter,
\be
\frac{2G_0M_B(R)}{c^2R} \leq \frac{8}{9}\left[1-\frac{9\pi G_0}{2c^4}\left.\left(\frac{m^2}{2}\psi ^2-\frac{1}{4}\eta \psi ^4\right)\right|_{r=R}R^2\right] \, .
\ee

%Sec.4.2.2%%%%%%%%%%%%%%%%%%%%%%%%%%%%%%%%%%%%%%%%%%%%%%
\subsubsection{The minimum mass of a bosonic object in generalized gravity theories} \label{sect4.2.2}

By assuming again that the function $\Delta(r)$ vanishes on the vacuum boundary, we can use Eq.~(\ref{minmass}) to estimate the minimum mass of bosonic objects in generalized gravity theories. Thus, we obtain
\be\label{minbos}
\frac{G_0M_{\rm eff}^{(\rm min)}}{c^2R}\geq \frac{2\pi G_0}{c^4}\left|\left(\frac{m^2}{2}\psi^2 - \frac{1}{4}\eta \psi^4\right)\right|_{r=R}R^2 \, ,
\ee
or, if the Higgs potential has a minimum at the surface,
\be
\frac{G_0M_{\rm eff}^{(\rm min)}}{c^2R} \geq \frac{\pi G_0}{2c^4}\frac{m^4}{\eta}R^2 \, .
\ee
Equation (\ref{minbos}) gives the following bound on the mean energy density, $\varepsilon _{\rm eff}^{(\rm min)} = \rho _{\rm eff}^{(\rm min)}c^2 = 3M_{\rm eff}^{(\rm min)}c^2/4\pi R^3$, of a bosonic objects with minimum mass,
\be
\varepsilon _{\rm eff}^{(\rm min)} \geq \frac{3}{2} \left|\left(\frac{m^2}{2}\psi^2 - \frac{1}{4}\eta \psi ^4\right)\right|_{r=R}=\frac{3}{8}\frac{m^4}{\eta} \, .
\ee
Since the total energy density consists of the sum of the energy densities of the baryonic matter and of the scalar field, $\varepsilon _{\rm eff}^{(\rm min)} = \varepsilon _{B}^{(\rm min)} + \varepsilon _{\psi}^{(\rm min)}$, we obtain the following constraint on the baryonic density energy,
\be
\varepsilon _{B}^{(\rm min)} \geq \frac{1}{2}\left|\left(\frac{m^2}{2}\psi^2 - \frac{1}{4}\eta \psi^4\right)\right|_{r=R}=\frac{1}{8}\frac{m^4}{\eta} \, .
\ee

By denoting the value of the potential at the surface of the compact bosonic object by
\be
Bc^2 = \left.\left(\frac{m^2}{2} - \frac{1}{4}\eta \psi^4\right)\right|_{r=R}c^2 = \left(\frac{m^4}{4\eta} \right)c^2 \, ,
\ee
we obtain the bound,
\be\label{114}
M_{\rm eff}\geq 2\pi BR^3 \, ,
\ee
yielding the minimum mass of any bosonic object. By assuming that the effective mass of the bosonic ``particle" is of the order of the proton mass, $M_{\rm eff} = 1.672 \times 10^{-24}$ g, and that its radius is of the order of the proton radius, $R = 0.875\times 10^{-13}$ cm, we obtain the value $B \approx 4\times 10^{14}$ g/cm$^3$. However, one may also obtain an estimate of the radius of a minimum-mass bosonic object from stability considerations.

We begin by defining the total energy $E$ (including the gravitational field contribution), corresponding to any compact object inside an equipotential surface $S$ of radius $R$, as \cite{LyKa88,GrJo92}
\begin{equation}
E = E_{M} + E_{F} = \frac{1}{8\pi}\xi _{s}\int_{S}[K]dS \, ,
\end{equation}
where $E_{M}=\int_{S}T_{\nu}^{\mu}\xi ^{\nu }\sqrt{-g}dS_{\mu}$ and $E_{F}$ are the energy of the matter and of the gravitational field, respectively. Here $\xi ^{\nu}$ is a Killing vector field of time translation, while $\xi _{s}$ is its value at $S$. $[K]$ denotes the jump across the shell of the trace of the extrinsic curvature of $S$, considered as embedded in the 2-space $t=\mathrm{const}$. $T_{\nu }^{\mu}$ is the energy-momentum tensor of the matter, as usual. This definition of the total energy is manifestly coordinate invariant.

By representing the metric on the surface of the compact object as $e^{-\lambda}=1-2G_0M_{\rm eff}/c^2R$, with the use of Eq.~(\ref{114}) it follows that  the total energy of a bosonic object with minimum mass inside radius $R$ can be written as
\bea
E &=& -\frac{c^4}{G_0}R\left[ 1-\left( 1-\frac{4\pi G_0 }{c^2}BR^{2}\right) ^{1/2}\right]
\nonumber\\
&\times &\left(1 - \frac{4\pi G_0}{c^2}BR^{2}\right)^{1/2} \, .
\eea
For a stable particle configuration, the energy must have a minimum, $\partial E/\partial R=0$, and this condition gives the following algebraic equation determining the radius $R$ as a function of $B$,
\bea\label{aleq}
&& \left(1-\sqrt{1-\frac{4 \pi  B G_0 R^2}{c^2}}\right) + \frac{4 \pi  B G_0 R^2}{c^2}
\nonumber\\
&&  \times \left(3 \sqrt{1-\frac{4 \pi  B G_0 R^2}{c^2}}-2\right) = 0 \, .
\eea
The solution of Eq. (\ref{aleq}) is
\begin{equation} \label{R_min_bosonic}
R = \frac{1}{6}\sqrt{\frac{11+\sqrt{13}}{2}}\frac{c}{\sqrt{\pi G_{0}B}} = r_{0}\frac{c}{\sqrt{\pi G_{0}B}} \, ,
\end{equation}
where $r_{0} = \sqrt{\left(11+\sqrt{13}\right)/2}/6 = 0.450$. Therefore, we can represent the lower bound giving the minimum possible mass of a bosonic object as
\begin{equation} \label{M_min_bosonic}
M_{\rm eff} \geq 2\frac{c^{3}}{G_{0}}\frac{r_{0}^{3}}{\sqrt{\pi G_{0}B}} \, .
\end{equation}

It is interesting to investigate whether the Chandrasekhar limit (\ref{Ch}) also applies to the {\it minimum} mass of a bosonic object, with the baryon mass substituted by an effective particle mass $m_{\rm qeff}$, representing the minimum mass of the particle composing the minimum mass system. If such a representation is possible, we must have
\begin{equation}\label{15}
\frac{m_{\rm Pl}^3}{m_{\rm qeff}^2} \sim \frac{c^{3}}{G_0}\frac{1}{\sqrt{\pi G_0B}} \, .
%\left(\frac{\hbar c}{G_0}m_{qeff}^{-4/3}\right)^{3/2}\sim \frac{c^{3}}{G_0}\frac{1}{\sqrt{\pi G_0B}}.
\end{equation}
Equation (\ref{15}) leads to the following expression of the effective mass of the ``elementary'' particle composing the minimum mass bosonic object,
\begin{equation}\label{16}
m_{\rm qeff} \sim \left( \frac{\hbar}{c}\right)^{3/4} B^{1/4} \, .
\end{equation}
Alternatively, in terms of the parameters of the Higgs potential we obtain
\be \label{Higgs*}
m_{\rm qeff} \sim \left(\frac{\hbar }{c}\right)^{3/4}\frac{m}{\eta^{1/4}} \, .
\ee
Thus, the effective particle mass forming a minimum mass bosonic object is determined {\it only} by physical constants associated with (non-gravitational) elementary particle physics. In particular, it is independent of the gravitational constant $G_0$. From its mathematical representation (\ref{15})-(\ref{16}), it follows that $m_{\rm qeff}$ must be relevant only when the system is quantum mechanical and involves high velocities and energies.

For $B = 4\times 10^{14}\;{\rm g/cm^{3}}$, we obtain the value $m_{\rm qeff} \sim 3.63\times 10^{-25}\;{\rm g}\approx 204\;{\rm MeV}$ for the minimum ``elementary" particle mass. For $B = 1.33\times 10^{15}\;{\rm g/cm^{3}}$
Eq. (\ref{16}) gives $m_{\rm qeff}\sim 4.9\times 10^{-25}\:{\rm g}\approx 275\;{\rm MeV}$. From an elementary particle physics point of view we can interpret the mass given by Eq. (\ref{16}) as the {\it minimum
mass of the stable quark bubble}, since it is of the same order of magnitude as the strange quark mass $m_{s}$ \cite{Olive}. Hence, the Chandrasekhar limit also applies also {\it composite} elementary particles, if we take $m_{\rm qeff}$ as representing the mass of the elementary constituent of the object. Moreover, the mass of the particle is generated by the effective value of the Higgs potential at the particle vacuum boundary, $B$. With respect to a scaling of the Higgs type potential of the form $B \rightarrow kB$, the effective minimum mass $m_{\rm qeff}$ scales as $m_{\rm qeff}\rightarrow k^{1/4}m_{\rm qeff}$.

%Sec.4.3%%%%%%%%%%%%%%%%%%%%%%%%%%%%%%%%%%%%%%%%%%%%%%%
\subsection{Dark energy and the general Higgs coupling} \label{sect4.3}

In the following section, we investigate the implications of our results for objects containing a significant amount of dark energy, assumed to be an ideal fluid satisfying the equation of state $P_{\rm DE} = w\rho_{\rm DE}c^{2}$.  Moreover, we consider that the ``matter'' inside the compact object consists of a scalar field, with mass density and pressure given by $\rho c^{2} =T^{t}_{t,S}$, $P_{r}=-T^{r}_{r,S}$, $P_{\theta}=-T^{\theta}_{\theta,S}$, $P_{\phi}=-T^{\phi}_{\phi,S}$, where $T^{\mu}_{\nu,S}$ denotes the energy-momentum tensor of the scalar field, whose non-trivial components are given by %(\ref{ente}),
\begin{eqnarray}
T^{t}_{t,S} &=& T^{\theta}_{\theta,S} = T^{\phi}_{\phi,S} = \frac{e^{-\lambda}}{2}\phi'^{2} + V \, ,
\nonumber\\
%\\
T^{r}_{r,S} &=& -\frac{e^{-\lambda}}{2}\phi'^{2} + V \, .
\end{eqnarray}
Note the asymmetric pressures, $P_{r} \neq P_{\theta}$, for the scalar field with radial profile. The conservation of the energy-momentum tensor $\nabla_{\mu}T^{\mu}_{\phantom{1}\nu} = 0$ yields
\begin{eqnarray}
\partial_{r}P_{\rm tot} &=& -(\rho_{\rm tot}c^{2} + P_{\rm tot})\frac{\nu'}{2} - (P_{r} - P_{\theta})\frac{2}{r} \, ,
\notag \\
&=& -(\rho_{\rm tot}c^{2} + P_{\rm tot})\frac{\nu'}{2} - (e^{-\lambda}\phi'^{2})\frac{2}{r} \, ,
\end{eqnarray}
where $\rho_{\rm tot} \equiv \rho + \rho_{\rm DE}$, $P_{\rm tot} \equiv P_{r} + P_{\rm DE}$.

For constant $\rho_{\rm DE}$, the gravitational field equation (\ref{00}) leads immediately to
\begin{equation}
e^{-\lambda} = 1-\frac{2G_0M(r)}{c^2r} - \frac{\Lambda r^{2}}{3}\equiv 1-2 \alpha(r) r^{2} \, ,
\end{equation}
where $\alpha (r)= G_0M(r)/c^2r^3+\Lambda / 6$, and the mass $M(r)$ is defined as the bare mass, without the dark energy contribution, i.e.
\begin{equation}
M(r) = 4\pi \int \rho(r) r^{2}~dr \, .  \label{Mdef}
\end{equation}
Hence, by setting $P_{\rm tot}(r=R) = w\rho_{\rm DE}c^{2}$, we again obtain lower and upper bounds on the mass-radius ratio, given by
\begin{equation}
u_{\pm} = \frac{2}{9}\left[1-\frac{3}{4}\left(1+w_{\rm DE}\right)\Lambda R^{2}\right] \pm \frac{2}{9}\sqrt{1+\frac{3}{4}w_{\rm DE}\Lambda R^{2}} \, ,  \label{olims}
\end{equation}
where
\begin{equation}\label{udef}
u \equiv \frac{G_0 M}{c^2R} \, .
\end{equation}
From the definition of mass $M(r)$ in Eq.~(\ref{Mdef}), the mass bounds can be translated into the bounds on the average density of the scalar sphere,
\begin{eqnarray}
\langle{\rho}_{\phi}\rangle &=& \langle\frac{e^{-\lambda}\phi'^{2}}{2}\rangle + \langle V(\phi) \rangle \, .
\end{eqnarray}
The average density is related to the total mass by $\langle\rho\rangle=3M/4\pi R^3$. Since $e^{-\lambda}\leq 1$ for $r\leq R$, we can perform integration by parts, giving
\begin{eqnarray}
\frac{M(R)}{4\pi}&\leq& \int_{0}^{R}\frac{\phi'^{2}}{2}r^{2}~dr+\int_{0}^{R}V~r^{2}~dr  \notag \\
&=& \int_{0}^{R}\left(V(\phi) -\frac{\phi V'(\phi)}{2} \right)~r^{2}~dr \, ,  \label{Poteqn}
\end{eqnarray}
where we have assumed $\phi(R)=0$, $\phi'(R)<\infty$ and used the flat-space equation of motion
\begin{equation}
\phi''(r)+\frac{2}{r}\phi'(r) = \partial_{\phi}V\equiv V'(\phi) \, .
\end{equation}

Lets us now consider the scalar potential, for example, for Higgs particle, which can be written as follows
\begin{equation}
\frac{\hbar^{2}}{c^{2}}V(\phi) = V_{0}+\frac{m^{2}}{2}\phi^{2}+g\phi^{3}+\lambda\phi^{4} \, .
\end{equation}
Then, by assuming that $\phi(r)$ is a decreasing function with respect to $r$, we may write
\begin{eqnarray}
V(\phi) - \frac{\phi V'(\phi)}{2}&=&\frac{c^{2}}{\hbar^{2}}\left( V_{0}-\frac{g}{2}\phi^{3}-\lambda \phi^{4}\right)
\notag\\
&<& \frac{c^{2}}{\hbar^{2}}\left[ V_{0}-\frac{g}{2}\phi(R)^{3}-\lambda \phi(R)^{4}\right]
\nonumber\\
&<& \frac{c^{2}}{\hbar^{2}}V_{0} \, ,  \label{Vpotb}
\end{eqnarray}
where we set $\phi(R)=0$. The bounds on $\langle\rho_{\phi}\rangle$ thus put constraints on the parameters $m$, $g$ and $\lambda$ of the scalar self-coupling. For the lower bound on the mass-radius ratio, using (\ref{olims}), (\ref{udef}), (\ref{Poteqn}) and (\ref{Vpotb}), we have
\begin{eqnarray}
\frac{c^2}{8\pi G_0}\frac{6}{R^2}u_{-}&\leq&\frac{c^{2}}{\hbar^{2}}V_{0} \, .
\end{eqnarray}
For $\Lambda R^{2}\ll 1$, this becomes
\begin{equation}
\frac{-\Lambda c^2}{8\pi G_0}\left(1+\frac{3}{2}w_{\rm DE} \right) \leq \frac{c^{2}}{\hbar^{2}}V_{0} \, .  \label{Vbound}
\end{equation}
Hence, a non-trivial bound only exists when $\Lambda >(<) 0$, $w<(>)-2/3$. For typical Standard Model~(SM) Higgs, the parameters $V_{0}$, $m$, $g$ and $\lambda$, at the tree level, are all related through the electroweak~(EW) symmetry breaking machanism, i.e.
\begin{eqnarray}
V_{0}=\frac{\mu^{2}v^{2}}{4} \, ,  \quad m^{2} = -2\mu^{2} = -8\frac{V_{0}}{v^{2}} \, ,
\nonumber\\
g = \frac{m^{2}}{2v} \, , \quad \lambda = \frac{m^{2}}{8v^{2}} \, ,  \label{VHiggs}
\end{eqnarray}
for the vacuum expectation value~(VEV) $v=246$ GeV, with $\mu^{2} < 0$. The troublesome fact that the vacuum energy $V_{0}$ is negative in the SM Higgs model remains an open problem in fundamental particle physics. (Though it is at least stable when all terms, $\phi^{2}$, $\phi^{3}$ and $\phi^{4}$ are positive at the tree level, the top quark contribution at the quantum level could nevertheless destabilize the potential at high energy scales; see~\cite{Bezrukov:2012sa,Degrassi:2012ry} for further details.) Our result, Eq. (\ref{Vbound}), simply demands that $V_{0}$ must at least match the dark energy density at the surface of a stable object,
\begin{eqnarray}
\rho_{\Lambda} = \frac{c^2\Lambda}{16\pi G_0} \leq \frac{c^{2}}{\hbar^{2}}V_{0} \, ,  \label{Vcond}
\end{eqnarray}
for $w_{DE}=-1$ and $\Lambda > 0$. Gravitational stability against the dark energy repulsion is satisfied by the Higgs particle provided that its zero-field value $V_{0}$ is normalized to satisfy the bound (\ref{Vcond}).  In standard EW symmetry breaking, instead of starting with the potential
\begin{equation}
V(\Phi) = \mu |\Phi|^{2}+\lambda |\Phi|^{4} \, ,
\end{equation}
we can always shift the ground state energy by adding constant term $\Delta V$ so that $V\to V+\Delta V$.  The value of the constant $V_{0}$ after the symmetry breaking is thus normalizable by the constant $\Delta V$.

%Sec.5%%%%%%%%%%%%%%%%%%%%%%%%%%%%%%%%%%%%%%%%%%%%%%%%
%%%%%%%%%%%%%%%%%%%%%%%%%%%%%%%%%%%%%%%%%%%%%%%%%%%%
\section{Mass and Poincar\'{e} stress bounds for electrically charged objects} \label{sect5}

The origin of the masses of charged elementary particles, in particular of the mass of the electron, is a problem that continues to attract the interest of physicists. The first attempts to explain the mass of the electron in purely electromagnetic terms go back to the early works of Abraham and Lorentz \cite{Jack,Rohr} who supposed that both momentum and energy are of a purely electromagnetic nature. Using the momentum conservation law, they inferred that, besides the external force acting on the electron, there must be a self-force given in terms of the particle charge density $\rho \left(\vec{r}, t\right)$ and current $\vec{j} \left(\vec{r}, t\right)$. The most serious defect of this model is related to the (im)possibility of having a highly localized charge density, which, in order to guarantee stability, is conditional on the presence of cohesive {\it non-electromagnetic} forces. This makes it impossible to formulate a purely electromagnetic mass model for matter, at least in non-gravitational theories.

Poincar\'{e} \cite{Poin} later modified the Abraham-Lorentz model, postulating the existence of non-electromagnetic forces, the so-called ``Poincar\'{e} self-stresses", which have to balance the electrostatic repulsion in order to guarantee the stability of charged particles, reducing the total force acting on the charge distribution to zero. He defined a symmetric non-electromagnetic tensor $P_{\nu}^{\mu}$, which has to be considered in addition to the symmetric electromagnetic energy-momentum tensor $T_{\nu}^{\mu}$, thus giving a total energy-momentum tensor $S_{\nu}^{\mu}=T_{\nu}^{\mu}+P_{\nu}^{\mu}$. The presence of $P_{\nu}^{\mu}$ should not modify the components of the electromagnetic momentum. In the particle's rest frame, the Poincar\'{e} self-stresses can be represented as $P_{\nu}^{\mu}={\rm diag}\left(\rho c^2,-p_r,-p_{\perp},-p_{\perp}\right)$, where $p_r$ and $p_{\perp}$ represent the equivalent radial and perpendicular pressures associated to the stresses. From a quantum theoretical point of view, Poincar\'{e} stresses were interpreted as a zero-point energy in \cite{Cas,Boyer, Man} and a new interpretation of the classical theory of electromagnetic mass was proposed in \cite{Sch}. Fermi's analysis of the contribution of the electromagnetic field to the inertial mass of the classical electron within special relativity was considered \cite{m5}, while the electromagnetic contributions to hadron masses were calculated, using the gauge/gravity duality, in \cite{m6}. With the development of general relativity, the construction of general relativistic electromagnetic mass models has also become an active field of research \cite{m1,m2,m3,m4}.

%Sec.5.1%%%%%%%%%%%%%%%%%%%%%%%%%%%%%%%%%%%%%%%%%%%%%%%
\subsection{Poincar\'{e} stress limits for charged objects} \label{sect5.1}

For a charged object in the presence of anisotropic stresses, the Einstein field equations take the form \cite{Bek}
\begin{equation}
-\frac{1}{r^{2}}\frac{d}{dr}\left( re^{-\lambda }\right) + \frac{1}{r^{2}} = 8\pi \rho + \mathcal{E}^2 \, ,
\label{00c}
\end{equation}
\begin{equation}
-e^{-\lambda }\left( \frac{\nu ^{\prime }}{r}+\frac{1}{r^{2}}\right) + \frac{1}{r^{2}}=-8\pi p_r + \mathcal{E}^2 \, ,
\label{rrc}
\end{equation}
\begin{eqnarray}\label{ththc}
-\frac{1}{2}e^{-\lambda }\left[ \nu ^{\prime \prime }+\frac{\nu ^{\prime 2}}{2}+\frac{\nu ^{\prime} - \lambda^{\prime }}{r}-\frac{\nu ^{\prime}\lambda ^{\prime }}{2}\right] =  -8\pi p_{\perp} - \mathcal{E}^2
%\, ,
\nonumber\\
\end{eqnarray}
\be
\frac{d}{dr}\left(r^2\mathcal{E}\right)=4\pi \sigma e^{\lambda /2}r^2 \, ,
\ee
where $\sigma$ is the electric charge density and $\mathcal{E} = |\vec{E}|$ is the electric field intensity. Defining the electric charge as
\be
Q(r)=4\pi \int_0^r{\sigma \left(r^{\prime}\right)e^{\lambda \left(r^{\prime}\right)/2}r^{\prime \;2}dr^{\prime}},
\ee
we obtain $\mathcal{E}(r)=Q(r)/r^2$. From Eqs.~(\ref{00c})-(\ref{ththc}) it follows that $\rho _{\rm eff} = \rho + \mathcal{E}^2/8\pi $, $p_{\rm eff}=p_r - \mathcal{E}^2/8\pi$, and $\Delta = p_{\perp} - p_r + \mathcal{E}^2/4\pi$, respectively. For the function $f(r)$, defined in Eqs.~(\ref{f(r)*}) and (\ref{fapprox}), respectively, we adopt the approximation $f(R)=(4/3)\pi \left(p_{\perp} - p_r + \mathcal{E}^2/4\pi\right)R^2$. We define the mass of the charged object as
\be
m_{\rm eff}(r) = m_{B}(r) + m_{\rm em}(r) \, ,
\ee
where
\be
m_B = 4\pi \int_0^r{\rho \left(r^{\prime}\right)r^{\prime \;2}dr^{\prime}} \, ,
\ee
is the baryonic component, and
\be
m_{\rm em}(r) = 4\pi \int_0^r{\frac{\mathcal{E}^2\left(r^{\prime}\right)}{8\pi}r^{\prime \;2}dr^{\prime}} \, ,
\ee
is the electromagnetic mass. With the use of the general equations (\ref{66a}) and (\ref{66b}), we obtain the following bounds yielding the maximum and minimum masses of a charged object in the presence of Poincar\'{e} stresses,
\bea
\frac{M_{\rm eff}}{R} &\leq& -\frac{1}{648} R^2 \left[\mathcal{E}^2+4 \pi  (P_{\perp}-P_{r})\right]
\notag \\
&\times& \Big[21 \mathcal{E}^2 R^{2} -  8 \left(21 \pi  P_{r} R^2+2\right)\Big]
\notag \\
&+& \frac{1}{12} R^2 \left(\mathcal{E}^{2}-8 \pi  P_{\perp}\right)+\frac{4}{9} \, ,
\eea
\bea\label{mmc}
\frac{M_{\rm eff}}{R} &\geq& 4 \pi  R^2 \left(P_{r}-\frac{\mathcal{E}^{2}}{8 \pi }\right)
\notag \\
&\times& \left[\frac{1}{3} \pi  R^2 \left(\frac{\mathcal{E}^{2}}{4 \pi }+P_{\perp}-P_{r}\right)-\frac{1}{2}\right] \, ,
\eea
where we have denoted $P_r=p_r(R)$ and $P_{\perp}=p_{\perp}(R)$, respectively.

Let us first consider the case $\Delta(r) = 0$, which requires $p_{\perp} = p_r + \mathcal{E}^2/4\pi$. Then, from Eq.~(\ref{66}), we obtain the following limit for the effective total mass-radius ratio,
\be
-2\pi \left(P_r-\frac{\mathcal{E}^2}{8\pi}\right)R^2\leq \frac{M_{\rm eff}}{R} \leq \frac{4}{9}\left[1-\frac{3\pi}{2}\left(P_r-\frac{\mathcal{E}^2}{8\pi}\right)R^2\right].
\ee
It is interesting to note that, even when the Poincar\'{e} stresses vanish, with $p_r=p_{\perp}=\rho =0$, there exist (purely electromagnetic) stable minimum and maximum mass limits, given by
\be
\frac{\mathcal{E}^2}{4}R^2\leq \frac{M_{e\rm m}}{R} \leq \frac{4}{9}\left(1+\frac{3\mathcal{E}^2}{16}R^2\right) \, .
\ee
In the presence of the non-electromagnetic components, the condition $P_r\leq \mathcal{E}^2/8\pi$ must be satisfied in order for a non-trivial minimum mass to exist. More generally, it follows that the surface value of the radial non-electromagnetic pressure must satisfy the constraints
\be\label{ppc}
 \frac{1}{2\pi}\left(\frac{\mathcal{E}^2}{4} - \frac{M_{\rm eff}}{R^3}\right) \leq P_r \leq \frac{2}{3\pi R^2}\Bigg[1-\frac{9}{4}\frac{M_{\rm eff}}{R}+\frac{3\mathcal{E}^2}{16}R^2\Bigg] \, .
\ee
Next, we consider the case in which $p_{r}=\mathcal{E}^2/8\pi$, $p_{\perp} \neq 0$ and $\Delta = p_{\perp}+\mathcal{E}^2/8\pi$. We then obtain the mass limits
\be
0 \leq \frac{M_{\rm eff}}{R} \leq \frac{1}{2}\left\{1-\frac{9}{\left[\left(8 \pi  P_{\perp}+\mathcal{E}^2\right) R^2+9\right]^2}\right\} \, ,
\ee
giving the following bound on the tangential pressure $p_{\perp}$,
\be
8\pi P_{\perp} \geq \frac{3}{R^2}\left(\frac{1}{\sqrt{1-2M_{\rm eff}/R}}-\frac{\mathcal{E}^2R^2}{3}-3\right) \, .
\ee

The radius of the charged compact object can be constrained from the assumption that its electrostatic energy $Q^2/R$ is of the same order of magnitude as its total mass-energy $M_{\rm eff}$. In this scenario, we obtain the general mass-charge relation $M_{\rm eff}=\left(1/\alpha _0\right)\mathcal{E}^2R^3$, where $\alpha _0$ is a constant. From Eq.~(\ref{mmc}), we then obtain the following lower bound yielding the minimum mass of a charged object in the presence of Poincar\'{e} stresses,
\begin{equation}
M_{\rm eff}\geq \frac{\mathcal{E}^{2}}{\alpha _{0}}\left\{ \frac{6\left(4-\alpha_{0}\right) \mathcal{E}^{2}+6\pi \alpha _{0}P_{r}}{\alpha _{0}\left[4\pi\left(P_{\bot}-P_{r}\right) + \mathcal{E}^{2}\right]
\left(\mathcal{E}^{2}-8\pi P_{r}\right) }\right\}^{3/2} \, .
\end{equation}
Equation (\ref{ppc}) gives the following constraint for the radial Poincar\'{e} pressure,
\be
P_r \geq \frac{\alpha _0-4}{8\pi \alpha _0}\mathcal{E}^2 \, .
\ee

%Sec.6%%%%%%%%%%%%%%%%%%%%%%%%%%%%%%%%%%%%%%%%%%%%%%%%
%%%%%%%%%%%%%%%%%%%%%%%%%%%%%%%%%%%%%%%%%%%%%%%%%%%%
\section{Implications of minimum mass limits for microscopic objects (fundamental particles)} \label{sect6}

In \cite{Wesson:2003qn} Wesson proposed the existence of two new fundamental mass scales, together with their corresponding lengths scales, derived from combinations of $\Lambda$, $\hbar$, $G$ and $c$. In this paper, we refer to these as the first and second Wesson mass (length) scales, given by
\begin{eqnarray} \label{Wesson}
m_{\rm W} &=& \frac{\hbar}{c}\sqrt{\frac{\Lambda}{3}} \, , \quad m_{\rm W}' = \frac{c^2}{G}\sqrt{\frac{3}{\Lambda}} \, ,
\nonumber\\
l_{\rm W} &=& \sqrt{\frac{3}{\Lambda}} \, , \quad \quad l_{\rm W}' = \frac{\hbar G}{c^3}\sqrt{\frac{\Lambda}{3}} \, ,
\end{eqnarray}
respectively. Originally, $m_{\rm W}$ was proposed as a fundamental minimum quantum of mass \cite{Wesson:2003qn}, though an alternative interpretation was suggested in \cite{min4}. The associated Compton wavelength $l_{\rm W}$ is of the order of the present day horizon size, which is equivalent to the length scale associated with the cosmological constant. By contrast, $m_{\rm W}'$ is of the order of the total mass of the present day Universe, approximately 70$\%$ of which is in the form of dark energy. The associated Compton scale $l_{\rm W}'$ is sub-Planckian, so that its physical meaning is unclear, though we include it in the definitions (\ref{Wesson}) for the sake of formal completeness. Interestingly, using the Wesson scales (\ref{Wesson}), the identity (\ref{Lambda_ident}) can be obtained in at least {\it three} different ways.

First, we note that setting $m_{\rm W}/R^3 \gtrsim \rho_{\Lambda}$, where $\rho_{\Lambda}$ is the minimum possible density of a gravitationally stable particle (in the presence of a positive cosmological constant) given in Eq. (\ref{dens_lim}), or, alternatively, $m_{\rm W}'/R^3 \lesssim \rho_{\rm Pl}$ implies
\begin{eqnarray} \label{rad_lim}
R \lesssim (\gtrsim) \ l_{\rm Pl}\left(\frac{m_{\rm Pl}}{m_{\rm W}}\right)^{1/3} = (l_{\rm Pl}^2l_{\rm W})^{1/3} \, ,
\end{eqnarray}
respectively. In other words, requiring the classical density of a fundamental mass quantum $m_{\rm W}$ to be greater than or equal to the minimum value given in (\ref{dens_lim}), or for the density of the Universe to be lower than the Planck density, yields the same scale, $R = (l_{\rm Pl}^2l_{\rm W})^{1/3}$, as either an upper or a lower bound on the radius of the system under consideration. Requiring the classical electron radius $r_e = e^2/m_e$ to satisfy {\it both} the lower and upper limits given in (\ref{rad_lim}) then yields
\begin{eqnarray} \label{r_e}
\frac{e^2}{m_e} \approx (l_{\rm Pl}^2l_{\rm W})^{1/3} \, ,
\end{eqnarray}
which is equivalent to (\ref{Lambda_ident}) up to numerical factors of order unity. Evaluating the left-hand side of (\ref{r_e}) gives $e^2/m_e = 2.98 \times 10^{-15}$ m, whereas evaluating the right-hand side using $\Lambda = 3.0 \times 10^{-56}$ cm$^{-2}$, the value of the cosmological constant inferred from observations \cite{Ostriker:1995rn,Tegmark:2003ud,Tegmark:2000qy,Hazra:2014hma,Zunckel:2008ti}, gives $(l_{\rm Pl}^2l_{\rm W})^{1/3} = 2.82 \times 10^{-15}$ m. Alternatively, comparing the left and right-hand sides of (\ref{r_e}) using only the observed values of the ``classical" constants $\left\{e,m_e,c,G,\hbar\right\}$ yields the estimate $\Lambda = 1.4 \times 10^{-56}$ cm$^{-2}$. This is strikingly close to the ``true" value, as first pointed out in \cite{min2,Funkhouser:2005hp}.

Second, an alternative derivation of Eq. (\ref{r_e}), based on minimising the total quantum uncertainty for a charged particle $-$ including canonical and gravitational contributions $-$ was given in \cite{min5}. This led to a ``cubic" MLUR of the form
\begin{eqnarray} \label{MLUR}
(\Delta x)_{\rm min} \approx (l_{\rm Pl}^2 \beta d)^{1/3} \, ,
\end{eqnarray}
where $\beta$ is a numerical constant (usually assumed to be of order unity \cite{Hossenfelder:2012jw,Ng:1993jb}) and $d$ denotes a distance being measured, or ``probed", with the aid of photon emission and absorption by a charged fundamental particle. The explicit inclusion of charge in the analysis presented in \cite{min5} offers a possible explanation for the appearance of the fine structure constant, $\alpha = e^2/(\hbar c)$, as a multiplicative factor on the right-hand side of (\ref{Lambda_ident}); this is the main difference between this relation and the form originally conjectured by Zel'dovich in \cite{Zel'dovich:1968zz,Zel1,Zel2}.

An MLUR of the form (\ref{MLUR}) was also proposed in \cite{Karolyhazy:1966zz,KFL}, in which it was argued that $(\Delta x)_{\rm min}$ represents a fundamental limitation to the accuracy of the measurement of the length of a geodesic, due to quantum gravity effects. However, (\ref{MLUR}) was not the first ``cubic" MLUR to be proposed in the context of fundamental limitations induced by quantum mechanical fluctuations of the gravitational field, or, equivalently, the spacetime metric. A similar but not identical relation,
\begin{eqnarray} \label{Bronstein-1}
(\Delta x)_{\rm min} \approx \left(\frac{\hbar c}{G \rho^2 V}\right)^{1/3} \, ,
\end{eqnarray}
was originally proposed by Bronstein in 1936 \cite{Bronstein}. Here, $\rho$ and $V$ denote the classical density and volume, respectively, of a quantum mechanical, self-gravitating ``particle". Hence, using $\rho \sim m/R^3$ and $V \sim R^3$, where $R$ denotes the classical radius, Eq. (\ref{Bronstein-1}) may be rewritten as \cite{min5}
\begin{eqnarray} \label{Bronstein-2}
(\Delta x)_{\rm min} \approx R\left(\frac{m_{\rm Pl}^2}{m^2}\right)^{1/3} \, .
\end{eqnarray}

The third ``derivation" of Eq. (\ref{Lambda_ident}), or equivalently (\ref{r_e}), follows from combining the MLUR (\ref{Bronstein-2}) with the existence of a minimum density $\rho_{\Lambda} = \Lambda c^2/(16\pi G)$, and of an effective mass $m_{\Lambda}$ and Compton wavelength $l_{\Lambda} = \hbar/(m_{\Lambda} c)$ for dark energy ``particles", such that $\rho_{\Lambda} \sim m_{\Lambda}/l_{\Lambda} ^3$. This yields \cite{min4}
\begin{eqnarray} \label{m_Lambda}
m_{\Lambda} \approx \sqrt{m_{\rm Pl}m_{\rm W}} \, , \quad  l_{\Lambda} \approx \sqrt{l_{\rm Pl}l_{\rm W}} \, .
\end{eqnarray}
Setting $m = m_{\Lambda}$ and $R = l_{\rm Pl}$ in (\ref{Bronstein-2}) then gives
\begin{eqnarray} \label{Bronstein-3}
(\Delta x)_{\rm min} \approx (l_{\rm Pl}^2l_{\rm W})^{1/3} \, ,
\end{eqnarray}
so that, applying this relation to the electron by setting $r_e \approx (\Delta x)_{\rm min}$ yields Eq. (\ref{r_e}). In other words, gravitationally stable minimum mass particles (i.e. those with mass $m_{\Lambda}$ and associated Compton wavelength $l_{\Lambda}$) have classical radius $l_{\rm Pl}$ but a minimum positional uncertainty of order $r_e = e^2/m_e$ according to Bronstein's relation. Furthermore, we note that this automatically ensures holography via \cite{min5}
\begin{eqnarray} \label{holography-1}
\left[\frac{(\Delta x)_{\rm min}}{l_{\rm W}}\right]^{3} = \frac{l_{\rm Pl}^2}{l_{\rm W}^2}
%\approx \frac{\hbar G \Lambda}{c^3}
\approx 10^{-120} \, .
\end{eqnarray}

In general, for $\beta \sim \mathcal{O}(1)$ and $R = l_{\rm Pl}$, Eqs. (\ref{MLUR}) and (\ref{Bronstein-2}) yield the same value of $(\Delta x)_{\rm min}$ when the effective gravitational mass associated with the length scale $d$, here denoted $m_d'$, takes Chandrasekhar form, i.e.
\begin{eqnarray} \label{d-Chand}
m_d' = \frac{c^2 d}{G} = \frac{m_{\rm Pl}^3}{m^2} \, .
\end{eqnarray}
Denoting $m_{d} = \hbar/(d c)$ as the effective quantum mechanical mass (i.e. ``Compton mass") associated with $d$, Eq. (\ref{d-Chand}) may be rewritten as $m = \sqrt{m_{\rm Pl}m_{d}}$. Setting $d = l_{\rm W} \approx 1/\sqrt{\Lambda}$ (its maximum possible value) and $m_{d} = m_{\rm W}$ (its minimum possible value) then gives $m = m_{\Lambda}$, which recovers Eq. (\ref{Bronstein-3}).

Interestingly, the factor $(m_{\rm Pl}^2/m_{\Lambda}^2)^{1/3}$ may also be expressed in terms of a new mass scale,
\begin{eqnarray} \label{Pandamass-1}
m_{\rm T} = (m_{\rm Pl}^2m_{\rm W})^{1/3} &=& (m_{\rm Pl}m_{\Lambda}^2)^{1/3}
\nonumber\\
&\approx& \left(\hbar^2\sqrt{\Lambda}/G\right)^{1/3} \, ,
\end{eqnarray}
as
\begin{eqnarray} \label{Pandamass-1}
\left(\frac{m_{\rm Pl}^2}{m_{\Lambda}^2}\right)^{1/3} = \frac{m_{\rm Pl}}{m_{\rm T}} \, .
\end{eqnarray}
Note that the mass $m_{\rm T}$ is {\it independent} of $c$ \cite{Burikham:2016rbj}. Based on a Generalized Uncertainty Principle of the form
\begin{eqnarray} \label{GUP1}
\Delta x \geq \frac{\hbar}{2\Delta p} + \beta \Delta p + l \, ,
\end{eqnarray}
a black hole with age comparable to the age of the Universe will stop radiating when its mass reaches the dual value $m_{\rm T}' = m_{\rm Pl}^2/m_{\rm T}$, at which point its Hawking temperature will be of order $T_{\rm H} \sim m_{\rm T}c^2/k_{\rm B}$. Holography persists for such remnant black holes, in arbitrary non-compact dimensions \cite{Burikham:2016rbj}. Finally, we note that, by Eq. (\ref{Lambda_ident}), $m_{\rm T}$ is related to the electron mass $m_e$ via
\begin{eqnarray} \label{m_Tm_e}
m_e = \alpha m_{\rm T} \, .
\end{eqnarray}
Using (\ref{r_e}), this is equivalent to the well-known relation
\begin{eqnarray} \label{lambda_er_e}
r_e = \alpha \lambda_e \, ,
\end{eqnarray}
where $\lambda_e = \hbar/(m_e c)$ is the electron's Compton wavelength. This relation may also be derived by modelling the electron as a gravitationally stable charged fluid sphere in canonical GR \cite{Bekenstein:1971ej} and is valid to first order in generalized theories including $\Lambda$CDM cosmology \cite{Boehmer:2007gq}.

The general considerations discussed above also have specific implications for the relationship between dark energy and the Higgs coupling, as considered in Sect. \ref{sect4.3}. Using the fact that $\rho_{\Lambda} \approx m_{\Lambda}/l_{\Lambda}^3 = m_{\Lambda}^4/(l_{\rm Pl}^3m_{\rm Pl}^3)$ and defining $V_0 \equiv m_{\rm H}^3/l_{\rm H} = m_{\rm H}^4/(l_{\rm Pl}m_{\rm Pl})$, the bound $\rho_{\Lambda} \leq (c^2/\hbar^2)V_0$ (\ref{Vcond}) may be written as
\begin{eqnarray} \label{Vcond-2}
m_{\rm H} \gtrsim m_{\Lambda} \, , \quad l_{\rm H} \lesssim l_{\Lambda} \, ,
\end{eqnarray}
i.e., to ensure gravitational stability, the Higgs mass must be greater than or equal to the effective mass of a dark energy particle. It is trivial to show that imposing $\rho_{\rm Pl} \geq (c^2/\hbar^2)V_0$ implies $m_{\rm H} \lesssim m_{\rm Pl}$ and $l_{\rm H} \gtrsim l_{\rm Pl}$. For the mass limits on bosonic objects obtained in Sect. \ref{sect4.2}, we find that requiring $\rho_{\rm Pl} \geq B \geq \rho_{\Lambda}$ implies the same bounds for $m_{\rm qeff}$. In terms of the parameters of the Higgs potential, the equivalent bounds on $m_{\rm qeff}$, as defined in Eq. (\ref{Higgs*}), then yield
\begin{eqnarray} \label{}
\frac{m^4}{m_{\rm Pl}^4} \lesssim \left(\frac{c}{\hbar}\right)^3\eta \lesssim \frac{m^4}{m_{\rm \Lambda}^4} \, .
\end{eqnarray}
For $m \approx m_{\Lambda}$, this gives
\begin{eqnarray} \label{}
10^{-120} \lesssim \left(\frac{c}{\hbar}\right)^3 \eta \lesssim 1 \, ,
\end{eqnarray}
whereas setting $m \approx m_{\rm Pl}$ implies
\begin{eqnarray} \label{}
1 \lesssim \left(\frac{c}{\hbar}\right)^3 \eta \lesssim 10^{120} \, .
\end{eqnarray}
Exploring the entire parameter range $m_{\Lambda} \leq m_{\rm qeff} \leq m_{\rm Pl}$, $m_{\Lambda} \leq m \leq m_{\rm Pl}$ therefore allows us to vary the Higgs field symmetry breaking parameter $(c/\hbar)^3\eta$ between its maximum and minimum possible values,
\begin{eqnarray} \label{}
10^{-120} \lesssim \left(\frac{c}{\hbar}\right)^3 \eta \lesssim 10^{120} \, .
\end{eqnarray}
Hence, the so-called ``cosmological constant problem", in which the na{\" i}ve calculation of the vacuum energy based on quantum field theory is of order $l_{\rm W}^2/l_{\rm Pl}^2 \approx c^{3}/(\hbar G\Lambda) \approx 10^{120}$ times larger than the measured value, is of vital importance in placing bounds on the parameters of the Higgs field in the presence of dark energy.

%Finally, before concluding this section, we note that, from the point of view of fundamental physics, the relation between the minimum mass of a compact bosonic object and its associated radius, given by Eqs. (\ref{M_min_bosonic}) and (\ref{R_min_bosonic}), respectively, and which we here denote by $M_{\rm min}$ and $R_{\rm min}$, is extremely intriguing Whilst, clearly, they obey the standard, gravitational, Schwarzschild type relation, $R_{\rm min} \approx GM_{\rm min}/c^2$, they are also related via
%\begin{eqnarray} \label{}
%R_{\rm min}M_{\rm min} \approx \frac{m_{\rm Pl}^2}{l_{\rm Pl}^2\rho_0} \, .
%\end{eqnarray}
%Therefore, when $\rho_0 \approx \rho_{\rm Pl}$, they also obey a Compton type relation, $R_{\rm min} \approx \hbar/(M_{\rm min} c)$. In other words, when the density of the bosonic object reaches the Planck density, its {\it maximum} possible radius will shrink to its Compton wavelength. Similar results hold for the maximum mass and radius of a quark star, defined in Eqs. (\ref{12*}).

%Sec.7%%%%%%%%%%%%%%%%%%%%%%%%%%%%%%%%%%%%%%%%%%%%%%%%
%%%%%%%%%%%%%%%%%%%%%%%%%%%%%%%%%%%%%%%%%%%%%%%%%%%%
\section{Discussions and final remarks} \label{sect7}

In the present paper, we have investigated the maximum and minimum mass limits for compact objects in generalized gravity theories, in which the total energy-momentum tensor can be expressed in the form  $T_{\mu \nu }^{(\rm tot)}=T_{\mu \nu }^{(\rm m)}+\theta _{\mu \nu }$, where $T_{\mu \nu }^{(\rm m)}$ denotes the ordinary matter energy-momentum tensor and $\theta _{\mu \nu }$ represents an additional contribution, coming from the generalization of the standard general relativistic model. A spatial variation of the gravitational coupling was also considered. The tensor $\theta _{\mu \nu }$ may be either purely ``physical" in origin, as considered in the example cases of scalar fields non-minimally coupled to gravity and of charged compact objects in canonical GR, or, alternatively, it may be interpreted as a geometric effect, due to the modification of the underlying gravitational theory.

As a first step in our study, we obtained the generalized TOV equation and Buchdahl inequalities, yielding general expressions for the upper and lower bounds on the mass-radius ratio of a stable compact object. We then used these results to study two particular cases of physical interest, namely, scalar-tensor theories with non-minimally coupled scalar fields and charged objects in canonical GR. For the scalar-tensor theories, we adopted a Higgs type potential for the self-interaction of the scalar field and assumed that this takes a non-vanishing but constant value at the vacuum boundary of the object, $B \neq 0$.

We found that the presence of a negative surface energy density implies the existence of a nonzero minimum mass and of a minimum density for a compact bosonic object, given by Eq.~(\ref{114}). In order to obtain an explicit representation of the minimum mass, rather than the minimum mass-radius ratio, we investigated the stability of minimum-mass objects using the condition of energy minimization to provide an alternative expression for the radius of the object. Using this procedure, the minimum mass may be expressed in terms of the gravitational constant and of the surface density $B$ only. Interestingly, the minimum mass also admits a Chandrasekhar type representation, given by Eqs.~(\ref{15})-(\ref{16}). In this representation, the minimum mass does not depend on the gravitational constant, and its numerical value is determined only by $\hbar$, $c$ and $B$. It is also interesting to note that, if $B$ is of the order of the nuclear density, the numerical value of the minimum mass coincides with the mass of the strange quark $s$ (in quantum chromodynamics it is usually assumed that the $u$ and $d$ quarks have negligible masses \cite{Olive}), to within an order of magnitude. In the case of the electron, with mass $m_e$, the surface density giving its mass, $B = \left(c/\hbar\right)^3m^4_e$,
%\be
%B = \left(\frac{c}{\hbar}\right)^3m^4_e \, ,
%\ee
is of order $B=15875.4$ g/cm$^3$, while for the proton $B=1.802\times 10^{17}$ g/cm$^3$.

 An important point, concerning the results obtained herein for bosonic objects, is their physical validity in light of various ``no go" theorems for static, localized scalar field configurations. In \cite{Bek1} and \cite{Bek2} it was shown that a static black hole cannot have {\it any} exterior classical scalar or massive vector fields. (See \cite{Heus4, n1,n2} for a detailed discussion of the no-hair theorems and of black holes with hair.) This result was obtained for a real scalar field $\psi $ with an energy-momentum tensor of the form $T_{\mu \nu}=\nabla _{\mu}\psi \nabla _{\nu}\psi -(1/2)g_{\mu \nu}\left(\nabla _{\alpha }\psi \nabla ^{\alpha}\psi +m^2\psi ^2\right)$, and follows from the vanishing of the integral $\int{\left(g_{\mu \nu}\nabla ^{\mu}\nabla ^{\nu}\psi+m^2\psi ^2\right)\sqrt{-g}d^4x}=0$, which requires $\psi$ to be identically zero throughout the {\it black hole exterior}. On the other hand, the necessary and sufficient conditions for the existence of a scalar soliton star were formulated in \cite{Lee1} and \cite{Lee2} as follows: i) the scalar field must be invariant under a space-independent phase transformation $\psi \rightarrow e^{i\theta }\psi $, and ii) in the absence of the gravitational field the theory must have non-topological soliton solutions. For mini soliton stars, the theory id required to satisfy only i), and not ii). From a physical point of view, satisfying condition i) implies the conservation of the generator of the phase transformation $N$, a condition which leads to a conserved particle number in the system. Since, from the beginning of our analysis, we have considered a {\it complex scalar field} which is invariant under a global phase transformation, condition i) is automatically satisfied by our models. Thus, the applicability of our results to at least some classes of boson stars, or mini soliton stars, is guaranteed by the phase invariance of the scalar field. However, if the scalar field is fundamental, in order to have a renormalizable theory, one should consider a second Hermitian scalar field $\chi$ \cite{Lee1}, with the potential having, for example, the degenerate vacuum form $U(\chi)=\left(m^2\chi ^2/2\right)\left(1-\chi /\chi _0\right)^2$, where $\chi=\chi _0$ gives the false (degenerate) vacuum state. The extension of our results to the two scalar field and two potential case will be considered elsewhere.

Furthermore, it is interesting to compare the mass limits for the scalar field stars and mini soliton stars, as obtained in \cite{Lee1} and \cite{Lee2}, to the results of the present study. The soliton contains an interior with $\chi \approx \chi _0={\rm constant}$, and a vacuum exterior. Since the scalar field is confined to the interior of the shell with radius $R$, it carries an energy $E_k\approx \pi N/R$, where $N$ is the conserved charge (the particle number). The shell also contains a surface energy $E_s=4\pi s R^2$, where the surface tension is $s=m\chi _0^2/6$. By minimizing the total energy $E=E_k+E_s$ we obtain $E_k=2E_s$, $M=12\pi sR^2$, $N=8sR^3$, and $M\sim N^{2/3}$, respectively \cite{Lee2}. If gravitation is included, the critical mass for the formation of a black hole can be estimated as $M_c\sim \left(48\pi G_0^2s\right)^{-1}\approx \left(l_{Pl}m\right)^{-4}m$, which for $m=30$ GeV gives numerical values of the order of $M_c\sim 10^{15}M_{\odot}$ and  $R\sim 10^2$ lightyears, respectively \cite{Lee1}. These values exceed by a large margin the global properties of the stellar type objects considered in the present paper. On the other hand, the radius of a mini soliton star is of order $R\sim 6\times 10^{-16}$ cm, and its mass is of order $m\sim 10^{10}$ kg, with a corresponding particle number $N\sim 10^{35}$ and a density $10^{41}$ times greater than the density of a neutron star \cite{Lee2}. These numerical values also exceed by many orders of magnitude the corresponding physical parameters of the bosonic type objects considered in our present analysis.

In the case of charged objects, we introduced anisotropic ``Poincar\'{e} stresses", needed to counterbalance electrostatic repulsion to ensure the stability of the object. After deriving maximum and minimum mass bounds for the Poincar\'{e} stress model, we used them to obtain constraints on the anisotropic stresses, modelled as a perfect anisotropic fluid. Thus, we obtained upper and lower bounds for both the radial and tangential components of the Poincar\'{e} stress tensor, expressed in terms of the charge and effective mass of the ``particle" (modelled as a microscopic fluid sphere).

The existence of an upper bound for the mass-radius ratio of stable compact objects also leads to upper bounds for other astrophysical quantities of major observational interest. One of these quantities is the surface red shift $z$,
which in a static spherically symmetric geometry can be defined generally as
\be
z = \left(1-\frac{2M_{\rm eff}}{R}\right)^{-1/2}-1 \, ,
\ee
where $M_{\rm eff}$ is the total effective mass of the compact object. For a general relativistic object satisfying the Buchdahl inequality $2M/R\leq 8/9$, we obtain the standard constraint on the gravitational redshift, $z \leq 2$. By contrast, with the use of Eq.~(\ref{58}), we obtain the following general restriction for the redshift in extended gravitational theories,
\be\label{z}
z\leq \frac{2\left[1+f(R)\right]}{1+4\pi w_{\rm eff}(R)} \, ,
\ee
where $w_{\rm eff}$ is the effective equation of state parameter for the matter and the function $f(R)$ can be approximated by Eq.~(\ref{fapprox}), so that $f(R)\propto \Delta R^2$. Therefore the function $f$ describes the effects of the anisotropic pressure distribution on the gravitational redshift, and also introduces a supplementary dependence of $z$ on the radius of the compact object.

As an astrophysical application of Eq.~(\ref{z}) we now consider the case of quark stars, in which quark matter is described by the MIT ``bag model", with equation of state $p_{{\rm eff}}=\left(\rho -4B\right)c^2/3$, where $B$ denotes the ``bag constant" \cite{quark}. Assuming that the surface density at the vacuum boundary of the star vanishes, $\rho \approx 0$, it follows that the quark star has negative effective pressure at its surface, $p_{\rm eff}(R) \approx -Bc^2 \leq 0$, where we have neglected a  numerical factor of the order of unity. Hence, we obtain the following constraint on the surface redshift,
\be
z \leq \frac{3}{\sqrt{1-12\pi \left(G_0/c^2\right)BR^2}} -1 \, ,
\ee
or, equivalently,
\be
z \leq \frac{3}{\sqrt{1-0.279\times \left(B/10^{14}\;{\rm g/cm^3}\right)\times \left(R/10^6\; {\rm cm}\right)^2}} - 1 \, .
\ee
For a compact star with surface pressure $B=10^{14}\;{\rm g/cm ^3}$ and radius $R=10$ km, we obtain $z \leq 2.533$. On the other hand, in the presence of a positive effective surface pressure, $p_{{\rm eff}}= Bc^2 \geq 0$, corresponding to a nonzero surface quark density of order $\rho (R)\approx 8B$, and again setting $R=10$ km, we obtain $z\leq 1.6526$. Though the latter bound is consistent with the surface redshifts obtained for objects obeying the standard Buchdahl bound (\ref{Buchdahl}), the former is not.

Finally, we considered the implications of the existence of minimum mass limits, in generalized gravity theories, for the stability of fundamental particles. Reviewing the existing literature, we found that several phenomenological approaches to quantum gravity $-$ involving minimum length uncertainty relations together with minimum mass bounds previously obtained for both charged and uncharged particles in the context of $\Lambda$CDM cosmology $-$  suggest a fundamental relation between dark energy and electro-weak scale physics (\ref{Lambda_ident}). Combining the classical mass bounds for bosonic objects obtained in Sect. \ref{sect4}, and the associated bounds on the Higgs parameters, with the simple assumption of the existence of a Compton wavelength, we were able to rewrite bounds on the symmetry breaking parameter $\eta$ in terms of the dimensionless constant $\hbar c\Lambda/G \approx 10^{120}$, which also characterizes the magnitude of the ``cosmological constant problem". This suggests a potential link between dark energy, the parameters of the Higgs field, and the gravitational stability of fundamental particles.

Thus, in this work, we have made the fundamental assumption that general relativity and other geometric theories of gravity can be extended, and remain valid, at the level of elementary particles, whose behavior is essentially quantum. The problem of the relevance of general relativity for understanding the structure and properties of elementary particle is a long standing and still unsolved problem in theoretical physics. One approach to this problem, which assumes that tensor fields play a fundamental role in the physics of strong interactions, was proposed in the framework of the so-called ``strong gravity" theory, introduced and developed in \cite{sg1,sg2,sg3, sg4}. This idea was formulated mathematically in a two-tensor theory of strong and gravitational interactions, where the strong tensor fields are governed by equations formally identical to the Einstein gravitational equations, apart from the coupling parameter $\kappa _f\approx 1 $ GeV$^{-1}$, which replaces the Newtonian gravitational coupling $k_g \approx  10^{-19}$ GeV$^{-1}$ \cite{sg3}. The equations for the strong field $f_{\mu \nu}$ and for the gravitational field $g_{\mu \nu}$ are obtained from the Lagrangian
\be
\mathcal{L} =\frac{1}{k_g^2}\sqrt{-g}R(g)+\frac{1}{k_f^2}\sqrt{-f}R(f)+\mathcal{L}_{fg}+\mathcal{L}_m \, ,
\ee
where the first term represents the standard general relativistic Lagrangian for the gravitational field,  while the second is its strong interaction analog, obtained by replacing $k_g$ by $k_f$ and $g_{\mu \nu}$ by $f_{\mu \nu}$. To give the elementary particles mass (as well as their weak gravitational interaction) a mixing term between the $f$ and $g$ fields is needed. A simple covariant mixing term was proposed in \cite{sg3}, and is given by
\bea
\mathcal{L}_{fg} &=& -\frac{M^2}{4k_f^2}\sqrt{-g}\left(f^{\mu \nu}-g^{\mu \nu}\right)\left(f^{\kappa \lambda}-g^{\kappa \lambda }\right)
\nonumber\\
&\times& \left(g_{\kappa \lambda }g_{\lambda \nu}-g_{\mu \nu}g_{\kappa \lambda}\right) \, .
\eea

In the limit in which the gravitational field may be ignored, $g_{\mu \nu} \rightarrow \eta _{\mu \nu}$, the gravitational equations of the strong gravity theory can be written as
\be
R_{\mu \nu}(f)-\frac{1}{2}f_{\mu \nu}R(f)=k_f^2T_{\mu \nu}^{(s)} \, ,
\ee
where
\be
k_f^2T_{\mu \nu}^{(s)}=\frac{1}{2}M^2\left(f^{\kappa \lambda}-\eta ^{\kappa \lambda }\right)\left(\eta _{\kappa \nu }\eta_{\lambda \nu}-\eta_{\mu \nu}g_{\kappa \lambda}\right)\frac{\sqrt{-\eta}}{\sqrt{-f}} \, .
\ee
Hence, the existence of maximum and minimum mass limits may be also considered in the framework of the strong gravity theory, which allows for the possibility of obtaining a systematic geometric description of both the gravitational and strong interaction properties of elementary particles.

However, we note that the current Standard Model theory of strong interactions, quantum chromodynamics (QCD), is based on the existence of conserved $SU(3)$ charge (color charge), whose existence has been experimentally confirmed. Though there is no (explicit) $SU(3)$ gauge symmetry in the strong gravity field equations, these are meant to describe only the gauge singlet sector of the strong interaction, mediated by massless and massive spin-2 particles coupled to the stress tensor, and {\it not} the sector including color charges. Hence, strong gravity is not expected to replace QCD, but to describe only certain aspects of strong interactions involving gauge singlet states {\it within} the canonical theory, using a gravitational type formalism. It is therefore justified to use strong gravity theory to explore the stability and confinement of gauge singlet mesons and baryons, though {\it not} the scattering that requires color charge interactions. Building on the formalism developed in the present work, we will investigate this problem in a future publication.

%Biblio%%%%%%%%%%%%%%%%%%%%%%%%%%%%%%%%%%%%%%%%%%%%%%%%
%%%%%%%%%%%%%%%%%%%%%%%%%%%%%%%%%%%%%%%%%%%%%%%%%%%%

\end{document}